\documentclass[fleqn,usenatbib]{mnras}

\usepackage{newtxtext,newtxmath}

\DeclareRobustCommand{\VAN}[3]{#2}
\let\VANthebibliography\thebibliography
\def\thebibliography{\DeclareRobustCommand{\VAN}[3]{##3}\VANthebibliography}

\usepackage{graphicx}	
\usepackage{amsmath}	
\usepackage[section]{placeins}
\usepackage{float}[H]
\usepackage[T1]{fontenc}
\usepackage{rotating}
\usepackage{fix-cm}
\usepackage{ulem}

\title[Molecular gas in the CMZ]{CHIMPS2: The physical properties and star formation efficiency of molecular gas in the Central Molecular Zone}

\author[S. M. King et al.]{
S.~M.~King$^{1}$\thanks{E-mail: arisking@ljmu.ac.uk (SMK)}, 
T.~J.~T.~Moore$^{1}$, 
S.~N.~Longmore$^{1}$, 
D.~J.~Eden$^{2,3}$, 
J.~D.~Henshaw$^{4}$, 
A.~J.~Rigby$^{5}$, 
R.~Rani$^{6}$
\\
$^{1}$ Astrophysics Research Institute, Liverpool John Moores University, IC2, Liverpool Science Park, 146 Brownlow Hill, Liverpool, L3 5RF, UK\\
$^{2}$ Department of Physics, University of Bath, Claverton Down, Bath, BA2 7AY, UK\\
$^{3}$ Armagh Observatory and Planetarium, College Hill, Armagh, BT61 9DB, UK\\
$^{4}$ Max-Planck-Institut für Astronomie, Königstuhl 17, 69117 Heidelberg, Germany\\
$^{5}$ School of Physics and Astronomy, University of Leeds, Leeds, LS2 9JT, UK\\
$^{6}$ Institute of Astronomy, National Tsing Hua University, General Building II No. 101, Sec. 2, Kuang-Fu
Road, Hsinchu 30013, Taiwan, R.O.C.\\
}

\date{Accepted XXX. Received YYY; in original form ZZZ}

\pubyear{2025}

\begin{document}
\label{firstpage}
\pagerange{\pageref{firstpage}--\pageref{lastpage}}
\maketitle

\begin{abstract}
We present Local Thermodynamic Equilibrium (LTE) estimates of the physical properties and star formation efficiency (SFE) of molecular gas in the Central Molecular Zone (CMZ), using new $^{12}$CO $J=2\to1$ observations from the James Clerk Maxwell Telescope. Combined with CHIMPS2 $^{12}$CO and $^{13}$CO $J=3\to2$, and SEDIGISM $^{13}$CO $J=2\to1$ data, we estimate a median excitation temperature of $T_{\rm ex} = 11$\,K for $^{13}$CO throughout the CMZ, with peaks exceeding 120\,K in the Sgr\,B1/B2 complex. Cooler gas dominates around Sgr\,A and nearby clouds. We derive a median H$_{2}$ column-density of $N(\mathrm{H}2) = 2 \times 10^{22}$\,cm$^{-2}$ and a total $^{13}$CO-traced gas mass of $M_{\rm gas} = 7 \times 10^6$\,M$_\odot$, consistent with previous estimates when accounting for spatial coverage. The instantaneous SFE is assessed using Hi-GAL compact sources detected at 70-$\micron$ and 160--500-$\micron$. The 70-$\micron$-bright SFE, tracing current star formation, is modest overall but elevated in Sgr\,B1/B2, the Arches cluster, and Sgr\,C. In contrast, the 160--500-$\micron$ SFE, tracing cold pre-stellar gas, is more broadly enhanced, particularly in the dust ridge clouds and towards negative longitudes surrounding Sgr\,C. The contrasting distributions suggest an evolutionary gradient in SFE, consistent with a transition from dense, cold gas to embedded protostars. Our results imply that the CMZ may be enter a more active phase of star formation, with large reservoirs of gas primed for future activity.
\end{abstract}


\section{Introduction}
\begin{figure*}
    \centering
    \includegraphics[width=2\columnwidth]{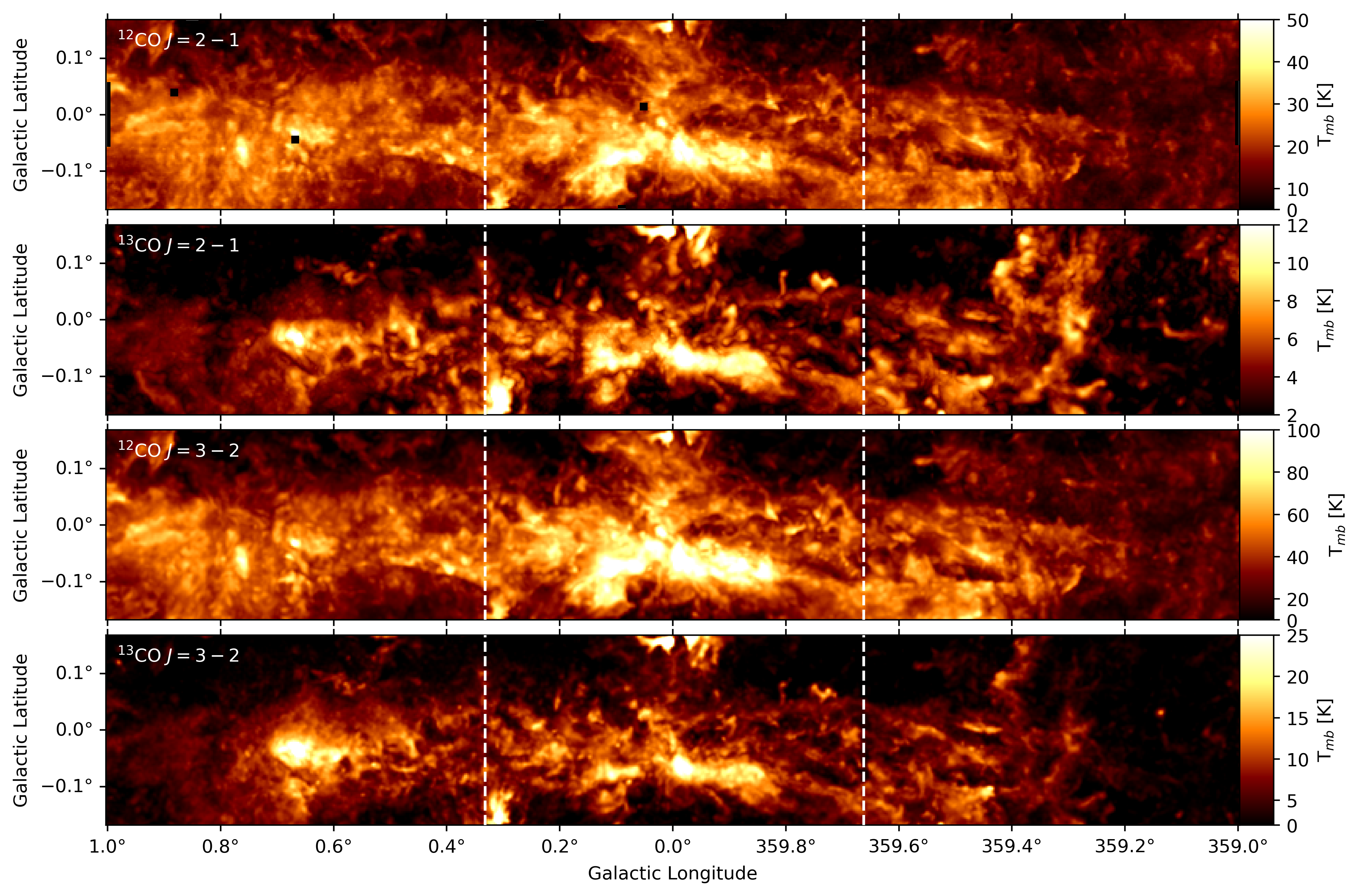}
    \caption{The peak main-beam temperature maps of the CMZ before smoothing for the transitions and isotopologues of CO.
    \textbf{Top}: $^{12}$CO $J = 2\to1$ (this work). The masked pixels (NaN values) are those with SNR $< 5$.\textbf{Second}: $^{13}$CO $J = 2\to1$ \citep{Schuller2017}. \textbf{Third}: $^{12}$CO $J = 3\to2$ \citep{Eden2020}. \textbf{Bottom}: $^{13}$CO $J = 3\to2$ \citep{King2024}. The top, third, and bottom maps were smoothed for the analysis to match the resolution of the second panel. The white dashed lines divide the regions containing one of the Sagittarius complexes (Sgr\,B, Sgr\,A, and Sgr\,C from left to right.)}
    \label{fig:13CO_comparison}
\end{figure*}

Star formation is one of the most important processes that drive galaxy evolution, enriching the interstellar medium (ISM) with metals and continually altering the balance between gas and stars within galaxies. Understanding the conditions and processes that govern star formation is crucial for explaining the evolutionary pathways of galaxies and for developing a predictive model of star formation. At present, it is not known how the star formation process is impacted by the physical conditions present in the local ISM.

The Central Molecular Zone (CMZ), the inner 500 parsecs of our Galaxy, contains approximately 5 per cent of the Milky Way's total molecular gas, estimated to be on the order of $10^{7}$\,M$_{\sun}$ \citep{Dahmen1998, Oka1998, Nakanishi2006, henshaw2023}. Despite its relatively small contribution to the total molecular gas budget, the CMZ hosts some of the most extreme environments for star formation in the Galaxy. Moreover, the CMZ contains a disproportionately large fraction of the denser, star-forming gas. Approximately 80 per cent of the Milky Way’s molecular gas at high densities, in which gravitational collapse is most likely to lead to star formation, is found here \citep{Morris&Serabyn1996, Ferrière2007, Longmore2013}.

Yet despite this abundance of dense molecular gas, the star formation rate per unit of dense gas in the CMZ is estimated to be an order of magnitude lower than in the spiral arms \citep[cf.][]{Lada2012,Longmore2013}, and understanding why the star formation efficiency (SFE) is so low in the CMZ, despite its high gas density, is a significant challenge \citep{Kruijssen2014, Rathborne2014, Barnes2017a, Dale2019}. 

Dense-gas temperatures derived from the observations of molecules such as metastable ammonia (NH$_3$) and para-formaldehyde (p-H$_2$CO) in the Galactic centre range from approximately 60\,K to over 100\,K in several regions, while highest gas temperatures are observed around Sgr\,B2, the 20-km\,s$^{-1}$ and 50-km\,s$^{-1}$ clouds, and in The Brick \citep{Ginsburg2016, Krieger2017}. 

These elevated temperatures suggest that the internal thermal energy of gas in the CMZ is higher compared to molecular clouds in the spiral arms. However, the broad line widths observed in dense gas within the central $\sim$250\,pc of the Galaxy are mostly due to turbulent motions rather than thermal broadening. Line widths here range between 5 and 10\,km\,s$^{-1}$ on parsec scales \citep{Henshaw2016b} and between 8 and 16\,km\,s$^{-1}$ on scales of individual molecular clouds ($\sim$10 to 50\,pc) \citep{Krieger2017}. Conversely, typical line widths of disc clouds are $3-5$ times smaller than in the CMZ. Typically, line widths here are $1-10$\,km\,$\mathrm{s}^{-1}$ \citep{Mills2017}. Analysis of molecular clouds and clumps in the Galactic Ring Survey (GRS) \citep{Jackson2006} also found that the typical line widths of molecular clouds in the Galactic plane are significantly narrower, often less than 10\,km\,s$^{-1}$ \citep{Rathborne2009}.

Previous studies, such as the investigation by \cite{Eden2015}, matched infrared-selected young stellar objects (YSOs) with dense molecular clumps across different Galactic environments to assess the SFE. Their findings demonstrated that the SFE, measured as the mean infrared luminosity-to-mass ratio ($L_\text{IR}$/$M_\text{clump}$), is slightly enhanced in spiral arms compared to interarm regions on kiloparsec scales. 

However, they also found that intrinsic variations within individual clumps dominate the overall range of SFE, with $L_\text{IR}$/$M_\text{clump}$ spanning three orders of magnitude and following a log-normal distribution. Additionally, a strong linear correlation was observed between $L_\text{IR}$ and $M_\text{clump}$, indicating that star formation activity scales directly with dense gas mass, independent of Galactic location. Our study builds on these findings by investigating whether similar trends and variations in SFE are observed in the CMZ, a unique Galactic environment characterized by extreme gas densities and turbulent conditions.

In a previous study \citep{King2024}, we presented the initial data for the $J = 3 \to 2$ transition of $^{13}$CO obtained from the CMZ as part of the CO Heterodyne Inner Milky Way Plane Survey 2 (CHIMPS2) \citep{Eden2020}.  Covering $358{\fdg}8 \leq \ell \leq 1{\fdg}2$ and $|b| \leq 0{\fdg}5$, our observations unveiled the complex structure of the molecular gas of the CMZ in improved detail. We estimated $^{13}$CO and H$_{2}$ column densities using simplified assumptions about temperature and explored the relationship between the integrated intensity of $^{13}$CO and the surface density of compact infrared sources identified by {\it Herschel} Hi-GAL \citep{Molinari2011, HIGAL2016, Elia2017}.

Building on this foundation, this paper presents a more in-depth analysis of the excitation conditions within the CMZ to extract accurate key physical properties such as excitation temperature, column density, and $^{13}$CO-traced gas mass. We use the same $^{12}$CO and $^{13}$CO $J = 3 \to 2$ data as previously \citep{Eden2020, King2024}, accompanied by $^{13}$CO $J = 2 \to 1$ from the SEDIGISM survey \citep{Schuller2017}. We also present new $^{12}$CO $J = 2 \to 1$ observations from the JCMT, which complete the isotopologue ladder for these transitions in the CMZ, allowing us to trace these properties by conducting a full excitation analysis under the assumption of Local Thermodynamic Equilibrium (LTE). We expand on previous studies by presenting column-density estimates using these derived excitation temperatures and total $^{13}$CO-traced gas mass estimates. 

Lastly, we investigate the instantaneous star formation efficiency (SFE) of the CMZ using the ratio of the integrated bolometric luminosity of Hi-GAL sources \citep{HIGAL2016, Elia2017, Elia2021} to gas mass. Section \ref{sec: Data and processing} details the data sets and processing methods used, Section \ref{sec: LTE analysis} discusses the LTE analysis of the data, Section \ref{sec: SFE} examines the star formation efficiency of the CMZ, and Section \ref{sec: Summary} summarises the findings of the article.

\vspace{-5mm}

\section{Data} \label{sec: Data and processing}
\subsection{The CHIMPS2 Survey}
The CHIMPS2 observations were carried out using the 15-m James Clark Maxwell Telescope (JCMT) in Hawaii, which has a diffraction-limited angular resolution of 15\,arcsec at 330.587\,GHz. The Heterodyne Array Receiver Program (HARP) was used in conjunction with the Auto-Correlation Spectral Imaging System (ACSIS) backend \citep{Buckle2009} to observe $^{12}$CO and $^{13}$CO in the CMZ with a binned channel width of 1\,km\,s$^{-1}$. The CMZ portion of the CHIMPS2 survey employed a systematic observing strategy, as outlined in \cite{Eden2020}, between Galactic longitudes $357^{\circ} \leq \ell \leq 5^\circ$ and latitudes $|b| \leq 0{\fdg}5$ in $^{12}$CO, and $358\fdg8 \leq \ell \leq 1\fdg4$, $|b| \leq 0{\fdg}5$ in $^{13}$CO. The spectral cubes have native angular resolutions of 14\,arcsec for $^{12}$CO and 15\,arcsec for $^{13}$CO. These were convolved with Gaussian beams of 9\,arcsec and 8\,arcsec, respectively, resulting in a final angular resolution of 17\,arcsec. The data have a spectral resolution of 1\,km\,s$^{-1}$ and an rms noise level in $T_{\rm A}^*$ of 0.58\,K for $^{12}$CO and 0.59\,K for $^{13}$CO, using 6 arcsec pixels.

\vspace{-0.5cm}

\subsection{The SEDIGISM Survey}
The SEDIGISM survey covers 78 square degrees of the inner Galaxy ($-60^{\circ} \leq \ell \leq 18^{\circ}$ $|b| \leq 0{\fdg}5)$ in the $J = 2 \to 1$ rotational transition of $^{13}$CO. The SEDIGISM data achieved a spatial resolution of 30\,arcseconds and a velocity resolution of 0.25\,km\,s$^{-1}$ \citep{Schuller2017}. The survey achieved mean rms sensitivities of $\sigma(T_{\rm{mb}}) \approx 0.8 \,\mathrm{K}$ per 0.25-km\,s$^{-1}$ velocity channel for $^{13}$CO, although the sensitivity varies across the survey region due to differing weather conditions. The main-beam efficiency of the APEX telescope used in the $^{13}$CO $J = 2 \to 1$ SEDIGISM data is $\eta_{\rm mb} = 0.73$ \citep{Gusten2006}.

\vspace{-0.5cm}

\subsection{Complementary JCMT Observations}
This paper also presents additional observations that complement the survey data sets, conducted using '\={U}'\={u}, one of the three receiver inserts within the Nāmakanui instrument on the JCMT \citep{Mizuno2020}. These data cover the $J = 2 \to 1$ rotational transition of $^{12}$CO at 230.538\,GHz within Galactic longitudes $359^{\circ} \leq \ell \leq 1{\fdg}0$ and latitudes $|b| \leq 0{\fdg}1$, as shown in the top panel of Fig.\,\ref{fig:13CO_comparison}. The observations have a native angular resolution of 24 arcsec at 230.538\,GHz, with a spectral resolution of 1\,km\,s$^{-1}$, and achieve an rms sensitivity of 1\,K per 0.25-km\,s$^{-1}$ channel on 10 arcsec pixels.

\subsection{Data Calibration and Resolution Matching} \label{subsec:data_calibration}
For consistency in resolution, the CHIMPS2 data were first smoothed using the Starlink {\sc kappa} routine {\sc gausmooth} \citep{Currie2014} to match the SEDIGISM spatial resolution of 30\,arcsec. The datasets were then converted from the corrected antenna-temperature scale ($T_{\rm A}^*$) to the main-beam brightness temperature scale ($T_{\rm mb}$) using:

\begin{equation}
T_{\rm mb} = \frac{T_{\rm A}^*}{\eta_{\rm mb}}.
\end{equation}

For this conversion, main-beam efficiencies of $\eta_{\rm mb} = 0.605$ and $0.72$ were adopted for the CHIMPS2 $J = 3 \to 2$ $^{12}$CO and $^{13}$CO data, respectively \citep{Buckle2009}.

Similarly, the JCMT $^{12}$CO $J = 2 \to 1$ observations were binned into 1-km\,s$^{-1}$ channels and smoothed to achieve a final angular resolution of 30\,arcsec, matching the SEDIGISM and CHIMPS2 datasets. The resulting rms sensitivity of the smoothed new data is 0.46\,K per 1-km\,s$^{-1}$ channel. The main-beam efficiency adopted for these data is $\eta_{\rm mb} = 0.57$ \citep{Mizuno2020}.

These processed datasets enable full LTE derivations of excitation conditions and physical parameters of CO-traced gas in the CMZ in the region covered by the JCMT $^{12}$CO $J = 2 \to 1$ observations where all four transitions are available. To ensure the reliability of extracted signals, all datasets were masked to remove emission with a signal-to-noise ratio (SNR) $<3$, corresponding to a 3$\sigma$ threshold. The data quality was assessed by calculating the fraction of sight lines with emission above 3$\sigma$ significance. Both the $^{12}$CO and $^{13}$CO transitions show nearly complete detection across the entire mapped region, with 100\% of sight lines exceeding 3$\sigma$ significance in the $J = 3\to2$ transitions and 99.95\% in the $J = 2\to1$ transitions.

\begin{figure}
    \includegraphics[width = \columnwidth]{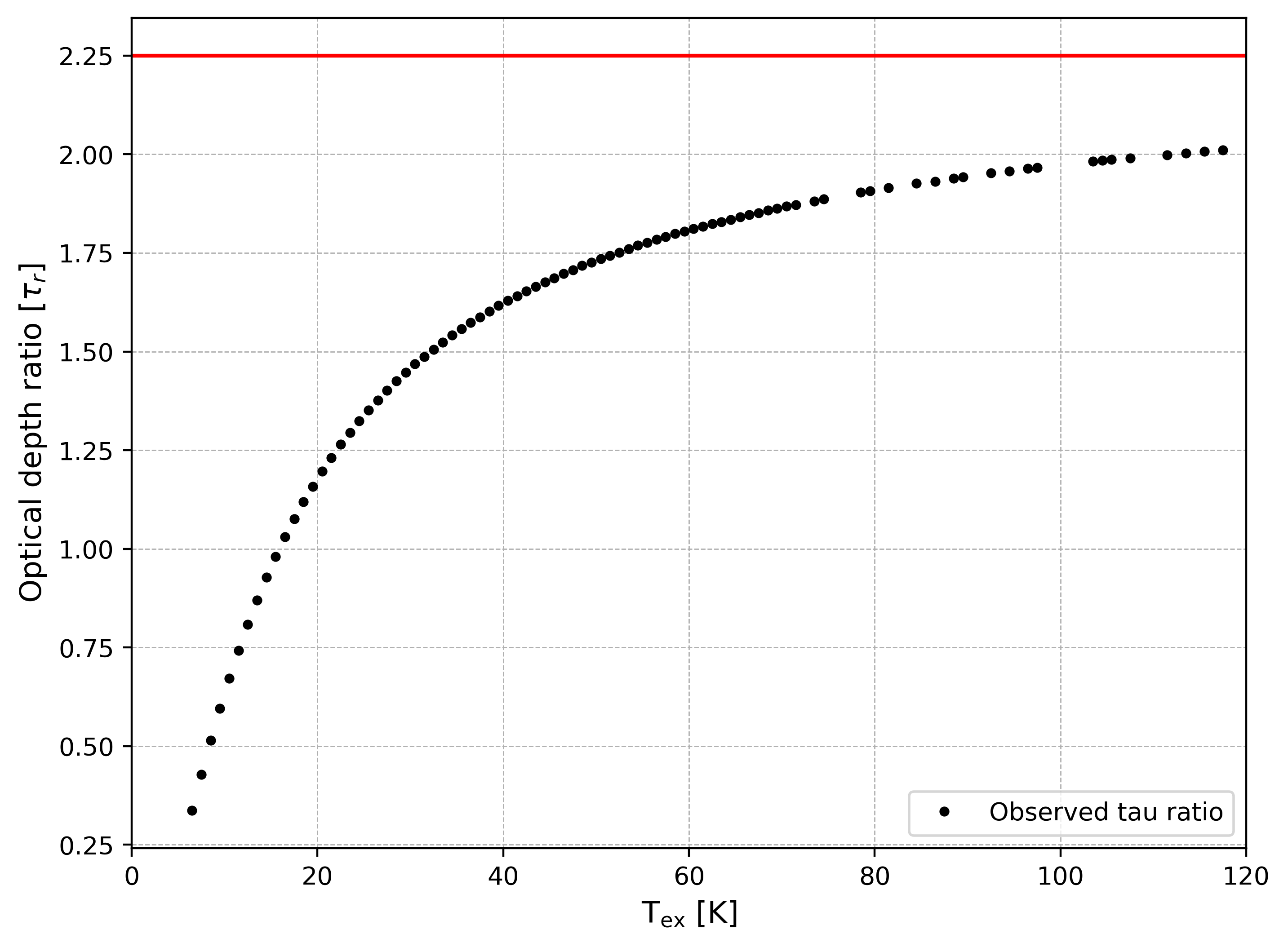}
    \caption{The  $J=3\to2/J=2\to1$ optical-depth ratio of $^{13}$CO as a function of excitation temperature ($T_{\rm ex}$) of $^{13}$CO determined using Equation\,\ref{Eqn: brents}. The red line represents the high-temperature limit at $\tau_r = 2.25$, above which no solution is found.}
    \label{fig:tau_ratio_hist}
\end{figure}

Finally, both the CHIMPS2 and SEDIGISM datasets were trimmed using {\sc wcsalign} to match the coverage and pixel grid of the new JCMT observations. The final rms sensitivities for these smoothed datasets are 0.2\,K for CHIMPS2 $J = 3 \to 2$ $^{12}$CO and 0.1\,K for $^{13}$CO, both lower than those reported in \citep{Eden2020} and \citep{King2024}. The trimmed and smoothed SEDIGISM $J = 2 \to 1$ $^{13}$CO data reached an rms sensitivity of 0.2\,K.

\vspace{-5mm}

\section{Physical Properties of the CMZ} \label{sec: LTE analysis}
In this section, we present estimates of the physical properties of the CMZ determined using $^{13}$CO under the assumption of local thermodynamic equilibrium (LTE). We assume that the brightness temperatures ($T_{\rm B}$) match the measured main-beam temperatures ($T_{\rm mb}$). Fig.\,\ref{fig:13CO_comparison} compares the peak main-beam temperatures for the transitions and isotopologues of CO used in this paper. The integrated-intensity maps were presented \citet{King2024}; however, in this work, we focus on peak-intensity maps to emphasize the maximum observed values, and to highlight regions with potentially higher excitation conditions. This also mitigates the impact of the very broad lines observed in the CMZ.

\begin{figure}
    \includegraphics[width = \columnwidth]{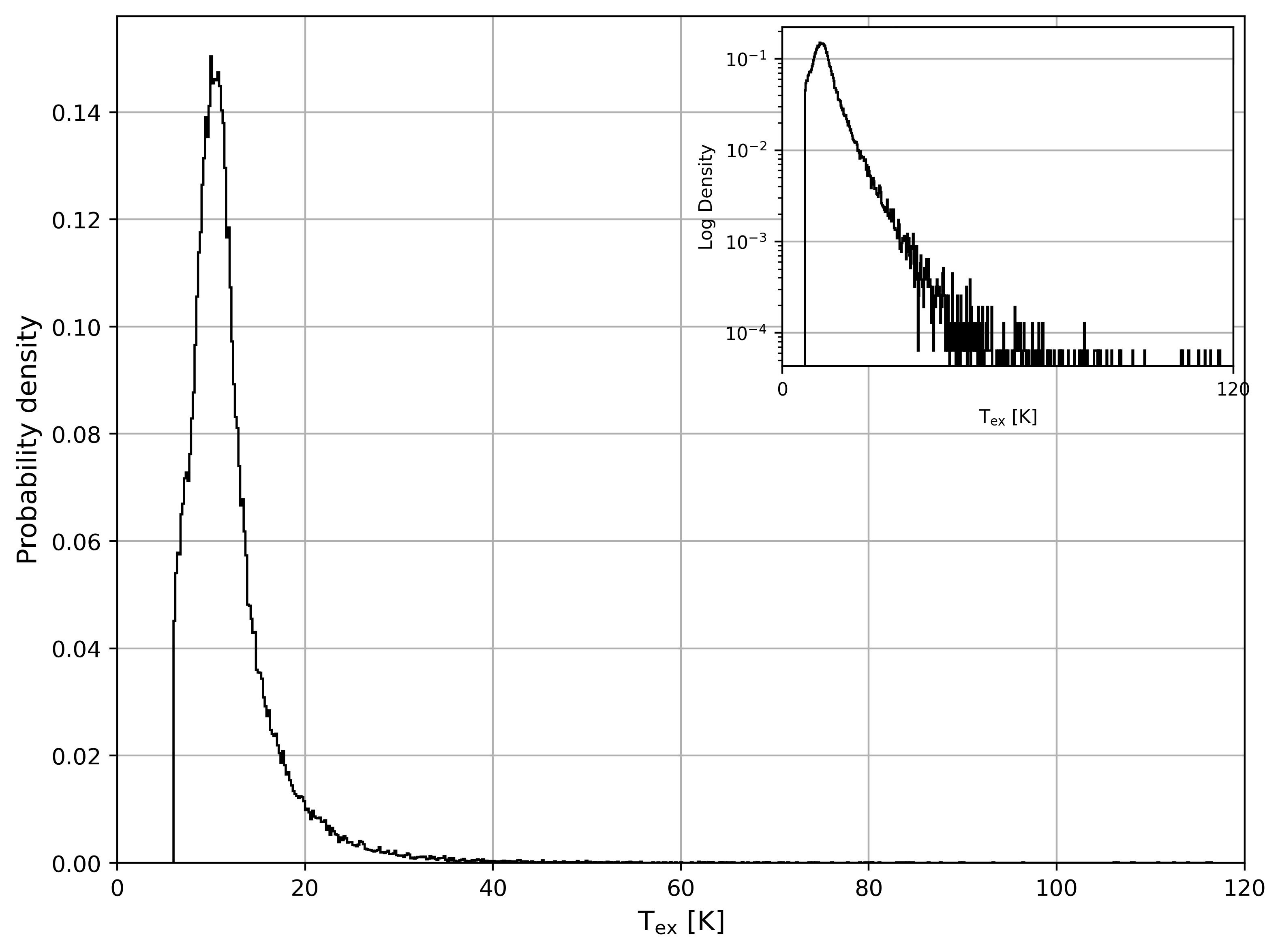}
    \caption{Histogram of the distribution of $^{13}$CO excitation temperature ($T_{\rm ex}$) values, calculated from the data under the assumption of Local Thermodynamic Equilibrium (LTE). The median $T_{\rm ex}$ is $11^{+2}_{-2}$\,K. The excitation temperature values were determined using the method described in Section~\ref{subsec: Temperature}. The inset plot shows the same distribution on a log-$y$ scale. }
    \label{fig:Tex_hist}
\end{figure}

\subsection{Optical depth ratio}
\subsubsection{Observed Ratio Calculation}
The $^{13}$CO optical depth of both transitions arises from the solution to the equation of radiative transfer, and was calculated using the ratio of $^{13}$CO to $^{12}$CO $T_{\rm mb}$ for each transition, by assuming that the $^{12}$CO optical depth ($\tau_{12}$) is much greater than 1, as follows:

\begin{equation}
    \label{Eqn: tau_ratio}
    \tau_{13} = -\ln \left(1 - (T_{\rm mb, 13} / T_{\rm mb, 12})\right)
\end{equation}  

Assuming Local Thermodynamic Equilibrium (LTE) and a single excitation temperature along the line of sight, the optical depth ratio of two adjacent transitions of the same isotopologue is derived from detailed balance, under the assumption that the population levels are described by a Boltzmann distribution, such that, for $J=3-2$ and $J=2-1$:

\begin{equation} \label{Eqn: brents}
\tau_{r} = \frac{\tau(3, 2)}{\tau(2, 1)} = \left( \frac{3}{2} \right) \left( \frac{1 - \exp\left( -\frac{6 h B}{k T_{\rm ex}} \right)}{\exp\left( \frac{4 h B}{k T_{\rm ex}} \right) - 1} \right)
\end{equation}

Where $B$ is the $^{13}$CO rotation constant in frequency units ($B = 55.10$ GHz), defined by $ B = h /8 \pi^2 \mathcal{I} $, $\mathcal{I}$ is the moment of inertia and $h$ is the Planck constant.

\begin{figure*}
    \centering
    \includegraphics[width=2\columnwidth]{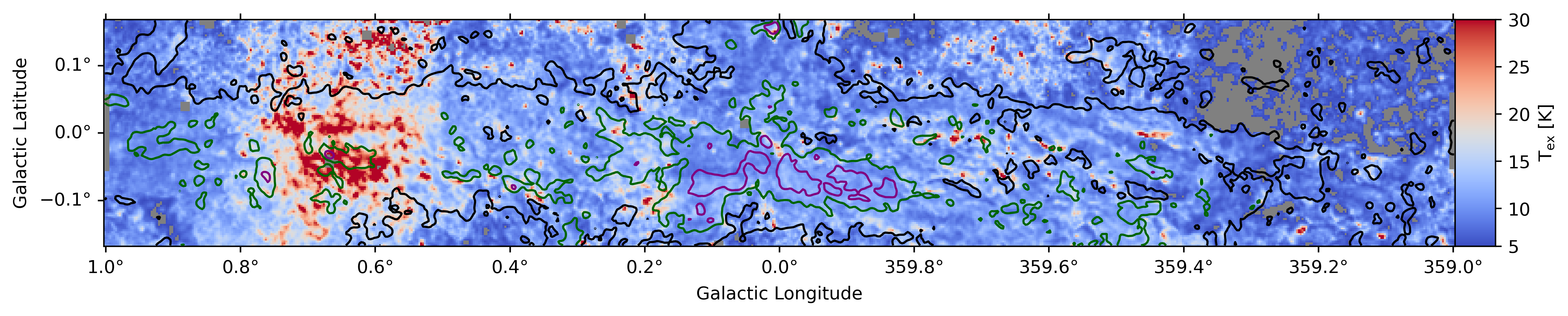}
    \caption{The distribution of the excitation temperature ($T_{\rm ex}$) across the Central Molecular Zone (CMZ), derived as in Section\,\ref{subsec: Temperature}. The distribution reaches its peak towards Sgr\,B, and is similar to the Hi-GAL dust temperature map shown in \citet{Molinari2011}. The contour shows the $^{12}$CO $J = 3 \to 2$ peak intensity from Fig.\,\ref{fig:13CO_comparison} at 30\,K (black), 60\,K (green) and 90\,K (purple). Grey pixels represent regions where $\tau_r > 2.25$, thus no valid solution to Equation\,\ref{Eqn: brents} is found.}
    \label{fig:Tex_map}
\end{figure*}

\subsection{Excitation Temperature Determination Using a Look-Up Table Method} \label{subsec: Temperature}
The $^{13}$CO excitation temperature ($T_{\rm ex}$) was determined using a computational look-up table approach based on Equation~\ref{Eqn: brents}. A theoretical look-up table of 87,362 optical depth ratios ($\tau_r$) was generated for excitation temperatures spanning 0–200\,K, this number of grid points was chosen to match the total number of pixels in the 2D maps providing a direct mapping between each ratio and its corresponding $T_{\rm ex}$.

For each pixel, the $^{13}$CO optical depths for the $J=3\to2$ and $J=2\to1$ transitions were computed using Equation~\ref{Eqn: tau_ratio}. The ratio of these optical depths was then compared to the theoretical values in the look-up table (as in Fig.\,\ref{fig:tau_ratio_hist}) to find the closest match. The associated $T_{\rm ex}$ with the closest matching theoretical $\tau_{\rm r}$ value was then assigned to that pixel to produce the distributions shown in Fig.\,\ref{fig:Tex_hist} and Fig.\,\ref{fig:Tex_map}.

\paragraph*{LTE limitations:} Equation\,\ref{Eqn: tau_ratio} will fail when $T_{\rm mb,13} \geq T_{\rm mb,12}$, or when $T_{\rm mb,12} = 0$, although the latter is not realistic given the 3-$\sigma$ threshold, we include it here for completeness. Typically, $T_{\rm mb,13}$ is lower than $T_{\rm mb,12}$, however, in regions with strong gradients in temperature along the line of sight, or in the presence of self-absorption, the assumption of a single excitation temperature can become inaccurate due to variations in optical depth allowing $T_{\rm mb,13}$ to approach or exceed $T_{\rm mb,12}$.

In regions of high optical depth, this can lead to systematic biases in the derived $\tau$-ratio. Since $^{12}$CO is typically more optically thick than $^{13}$CO, its emission will be dominated by foreground layers, while $^{13}$CO can trace deeper, potentially cooler gas. These effects are more likely to occur in regions better traced by $^{13}$CO, specifically where strong emission is observed in both transitions and where high optical depths or $^{12}$CO self-absorption are more prevalent. Furthermore, when $\tau_{13, 2-1}$ is small and $\tau_{13, 3-2} \approx 0$, Equation\,\ref{Eqn: brents} will also be invalid. This typically occurs in regions where $^{13}$CO is not sufficiently detected, which again is unlikely in this case given the 3-$\sigma$ threshold, and is included here only for completeness.

An additional limitation arises from the maximum allowable $\tau$-ratio under the assumption of LTE and a constant $T_{\rm ex}$ along the line of sight. Under these conditions, the ratio $\tau_r$ is capped at a theoretical maximum of $9/4$ shown as the red line in Fig. \ref{fig:tau_ratio_hist}. This arises because the bracketed exponential term in Equation \ref{Eqn: brents} approaches $3/2$ for $T_{\rm ex} \gg hB/k$, causing the equation to simplify to the general form $\tau_r = (J+1)/J)^2$ (where J = 2 is the common rotational level). The limit is a direct consequence of the Boltzmann distribution, which governs the population of energy levels in thermal equilibrium. 

The assumption of LTE also imposes limitations. The transitions used in this analysis have different critical densities, with the $J=3\to2$ transition requiring higher densities for thermalisation than $J=2\to1$. If the gas is sub-thermally excited, particularly for the higher transition, collisional excitation will be inefficient and the observed $T_{\rm mb}$ will be lower than expected under LTE assumptions. This would lead to a systematic underestimation of $T_{\rm ex}$.

The LTE assumption and the transitions used in this work also suggest a lower cut off of 5.5\,K for the derived excitation temperatures, being approximately the excitation energy of the $J=1$ level of CO.  At this temperature, almost all molecules are in either the $J=1$ or $J=0$ levels and the ($J=3\to2$)/($J=2\to1$) line ratio becomes an unreliable thermometer. The system also enters a regime where the excitation temperature converges to values close to that of the Cosmic Microwave Background (2.73\,K), causing the ratio to lose its diagnostic power and become dominated by uncertainties. 

Lastly, noise-affected pixels can return unreliable results, particularly in low-intensity regions where fluctuations in $T_{\rm mb,12}$ and $T_{\rm mb,13}$ may be comparable to the signal itself. Nevertheless, deriving the excitation temperature from optical-depth ratios remains a robust and physically well-motivated method.

\vspace{-5mm}

\subsection{Column Density and Total gas mass estimates} \label{subsec: column_density_method}
With an estimate of the $^{13}$CO excitation temperature, and assuming the high-temperature approximation for the partition function $Z \approx kT_{\rm ex} / hB$, we can then determine the column density of $^{13}$CO using the following equation:

\begin{multline} \label{Eqn: col_dense_13}
N({\rm{^{13}CO}})=\frac{8 \pi}{7 c^2}\frac{k T_{\rm ex}}{hB}\frac{\nu_{32}^{2}}{A_{32}}e^{\frac{6hB}{kT_{\rm ex}}}\\\left(1-e^{-\frac{h\nu_{32}}{kT_{\rm ex}}}\right)^{-1} \int \tau_{\rm 13}\,d\nu \quad \text{cm}^{-2}
\end{multline}

Where $T_{\rm ex}$ is the excitation temperature of $^{13}$CO, which is assumed to be the same for both the $J= 3 \to 2$ and $J = 2 \to 1$ transitions (i.e., both transitions are thermalised), determined in Section \ref{subsec: Temperature}. $\nu_{32}$ and $\tau(\nu)$ are the transition frequency and the $^{13}$CO optical depth of the 3--2 line, respectively. The integration is performed over all velocity channels between $v = \pm 200$\,km\,s$^{-1}$, assuming that the excitation temperature remains uniform across all velocity components, meaning no line-of-sight temperature gradients are present. 

We then used the $^{13}$CO column density to estimate the H$_2$ column-density, $N(\mathrm{H}_2)$ assuming an appropriate abundance ratio for $^{13}$CO to H$_{2}$, $^{13}\mathrm{CO} / \mathrm{H}_2$ (see subsection \ref{subsec: column_density}). The H$_{2}$ column density is therefore:

\begin{equation} \label{Eqn:H2 column density}
N(\mathrm{H}_2) = \frac{N({}^{13}\mathrm{CO})}{\left[^{13}\mathrm{CO} / \mathrm{H}_2\right]}
\end{equation}

When summed over the observed area, $N(\mathrm{H}_2)$ gives the total mass of gas traced by $^{13}$CO, in a similar manner to that described in appendix A of \cite{Kauffmann2008}, given by:

\begin{equation} \label{Eqn: H2_mass_equation}
M_{\rm tot} = \mu_{\rm H_{2}}\, m_{\rm H}\, A_{\rm proj} \sum \,N({\rm H_{2}})
\end{equation}

where $\mu_{\rm H_{2}} \approx 2.72$ is the mean molecular weight per H$_{2}$ molecule, accounting for a helium fraction of 0.25, $m_{\rm H}$ is the atomic hydrogen mass in kg, and $A_{\rm proj}$ is the physical area subtended by a single pixel on the plane of the sky, calculated as $A_{\rm proj} = d^2 \cdot A_{\rm pix}$. Here, $d$ is the distance to the source (8.178\,kpc \citep{gravity2019}) in centimetres, and $A_{\rm pix} = \Delta b \cdot \Delta l$ is the angular size of the pixel in steradians, where $\Delta b$ and $\Delta l$ are the pixel dimensions in Galactic latitude and longitude, respectively, expressed in radians. Finally, $N({\rm H_{2}})$ is the column density determined using Equation~\ref{Eqn:H2 column density}.

\subsection{Results}
\subsubsection{Excitation Temperature}
\begin{figure*}
    \includegraphics[width=2\columnwidth]{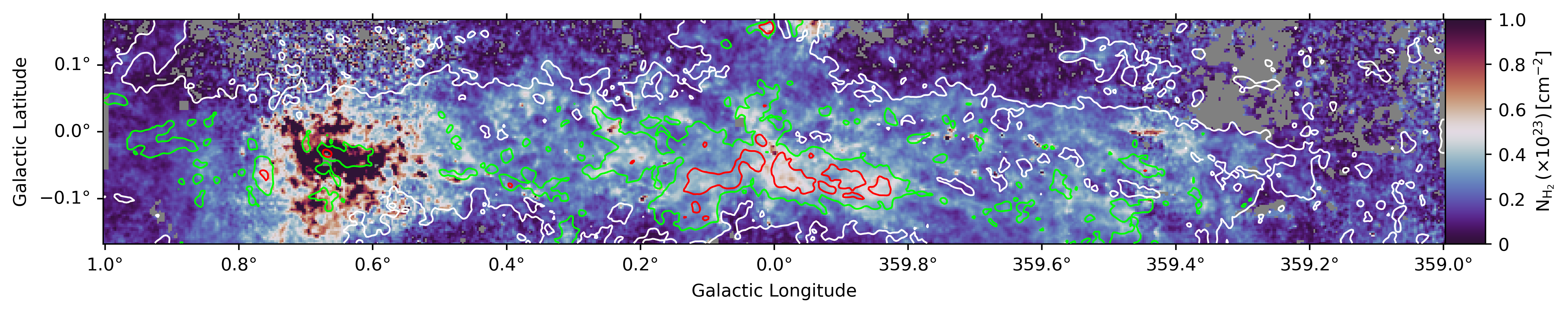}
    \caption{The H$_{2}$ column density across the CMZ. The peak column density of $2 \times 10^{25}\mathrm{cm}^{-2}$ is observed towards the Sgr\,B1/B2 complex (G0.070-0.059). The contour shows the $^{12}$CO $J = 3 \to 2$ peak intensity at $T_{\rm mb} = 30$\,K (white), $60$\,K(dark-grey), and $90$\,K(red).}
    \label{fig:H2_column_density}
\end{figure*}

The calculations using Equations\,\ref{Eqn: tau_ratio} and \ref{Eqn: brents} have produced a $^{13}$CO LTE excitation temperature for each pixel in Fig.\,\ref{fig:13CO_comparison} for which a solution is possible. Fig.\,\ref{fig:Tex_hist} shows the distribution of these $T_{\rm ex}$ values as a histogram, while Fig.\,\ref{fig:Tex_map} shows the spatial distribution extending from 5.5\,K to 30\,K.

The upper limit for temperatures determined using this method is approximately $200$\,K, as Equation~\ref{Eqn: brents} becomes increasingly insensitive and subject to noise fluctuations in $\tau_r$ at high temperatures. The optical-depth ratio $\tau_{32}/\tau_{21}$ is less sensitive to changes in excitation temperature above $\sim50$\,K (see Fig.\,\ref{fig:tau_ratio_hist}). The grey pixels in Fig.\,\ref{fig:Tex_map} represent positions where no valid solution could be obtained due to missing or unreliable optical depth values in one or both transitions. 

In bright, compact regions such as Sgr\,B, the $\tau_{13}$ calculation can become less accurate as high optical depths may saturate the line ratios, while calibration uncertainties and differential beam filling between the main line and isotopologue become increasingly significant at higher intensities.

The mean excitation temperature across all pixels is $12 \pm 1$\,K. Here, the uncertainty represents the mean of the per-pixel uncertainties, derived via quadrature error propagation from $\tau_{\rm r}$. The median is $11^{+2}_{-2}$K, where the uncertainties give the range covering the first and third quartiles. The highest temperatures in the CMZ are seen around the Sgr\,B1/B2 (G0.070-0.059) complex, where values reach $T_{\rm ex} > 30$\,K, the highest temperature is also found here at $T_{\rm ex} > 120$\,K. However, we find mostly cool temperatures across much of the CMZ, with isolated hot regions, e.g.\ at $\ell = 0\fdg2 ,b =-0\fdg1$, and $359\fdg8 \leq \ell \leq 359\fdg6, b =0^{\circ}$, while temperatures in Sgr\,A appear far cooler, with little distinction between the individual Sgr\,A clouds.

\vspace{-0.5cm}

\subsubsection{Column Density} \label {subsec: column_density}
Using the derived excitation temperatures seen in Fig.\,\ref{fig:Tex_map}, we estimate the mean $^{13}$CO column density to be $N(^{13}{\rm CO}) = (3.0 \pm 0.5) \times 10^{16}$\,cm$^{-2}$, where the uncertainty reflects the average per-pixel error propagated from uncertainties in $T_{\rm ex}$. The median is $N(^{13}{\rm CO}) = (3^{+1}_{-1}) \times 10^{16}$\,cm$^{-2}$, the range covering the first and third quartiles. 

Assuming a $^{13}$CO-to-H$_{2}$ abundance ratio of $^{13}\mathrm{CO} / \mathrm{H}_2 = 1.29\times10^{-6}$, we estimate the H$_{2}$ column-density using Equation\,\ref{Eqn:H2 column density}. Fig.\,\ref{fig:H2_column_density} shows the resulting spatial distribution of the H$_{2}$ column density. The mean observed column density is $N({\rm{H_{2}}}) = (3 \pm 0.2) \times 10^{22}$\,cm$^{-2}$, while the median is $(2^{+0.9}_{-0.7}) \times 10^{22}$\,cm$^{-2}$, again quoting the first and third quartiles. The peak value is $2 \times 10^{25}$\,cm$^{-2}$, located within Sgr\,B2.

Overall, the observed column-densities are consistent with previous CO-based \citep{nagai2007} and dust-continuum studies \citep{marsh2017, Battersby2025, tang2021}, which report similar H$_2$ values for the CMZ. A forthcoming paper will explore the relationship between dust and gas in the CMZ.

\subsubsection{Uncertainty Estimation}
To quantify the uncertainties in the derived $T_{\rm{ex}}$ and H$_2$ column density, we employed a Monte Carlo propagation approach. The pixel-by-pixel uncertainties in the CO line ratios were first estimated using the measured rms maps of the individual CO datasets and standard error propagation. These ratio uncertainties were then propagated to the optical depth ratio, which serves as the primary input for the excitation temperature calculation.

For each valid pixel, 1000 Monte Carlo realisations of the $T_{\rm ex}$ map were generated by perturbing the line ratios with Gaussian noise according to their propagated uncertainties. The excitation temperature for each realisation was computed using the same method described in section\,\ref{subsec: Temperature}. The final $T_{\rm ex}$ uncertainty map was obtained as the standard deviation across all realizations, yielding a typical 1$\sigma$ uncertainty of $\sim0.58$~K.

The $T_{\rm ex}$ realisations were subsequently propagated to the $^{13}$CO column density using Equation\,\ref{Eqn: col_dense_13}.  Each realisation of $N(^{13}$CO) was converted to an H$_2$ column density using the same assumed abundance ratio as above. The H$_2$ uncertainty map was computed as the standard deviation over all Monte Carlo realisations. The median uncertainty across the map is $5.7\times10^{21}$\,cm$^{-2}$, corresponding to a typical fractional uncertainty of $\sim30\%$ relative to the median H$_{2}$ column density of $2\times10^{22}$\,cm$^{-2}$.

\subsubsection{Total Molecular Gas Mass}

\begin{figure}
    \includegraphics[width = \columnwidth]{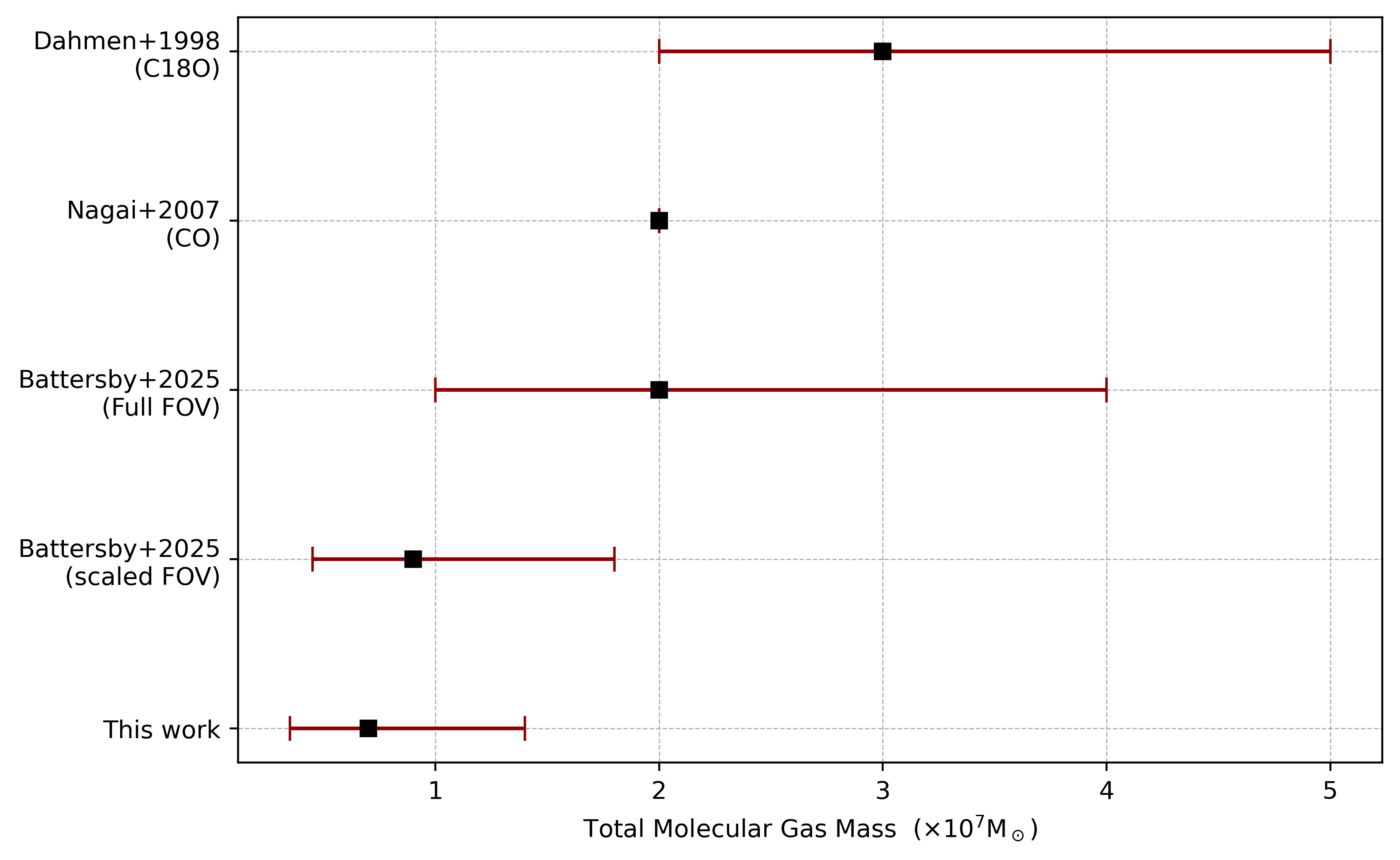}
    \caption{Total molecular gas mass estimates for the CMZ from this work and previous studies. The \citet{Battersby2025} mass estimate, originally derived from dust continuum over a larger region, has been scaled to the same effective field of view as our LTE analysis to facilitate a consistent comparison. Uncertainties for \citet{nagai2007} are not reported in the original study and are therefore omitted here.}
    \label{fig:mass_comp}
\end{figure}

\begin{figure*}
    \centering
    \includegraphics[width=2\columnwidth]{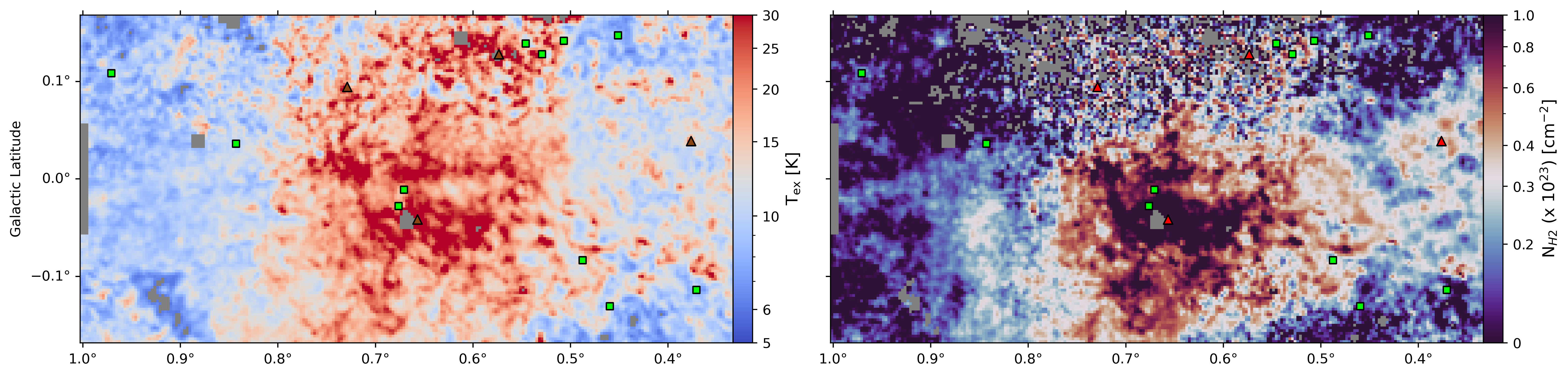}
    \includegraphics[width=2\columnwidth]{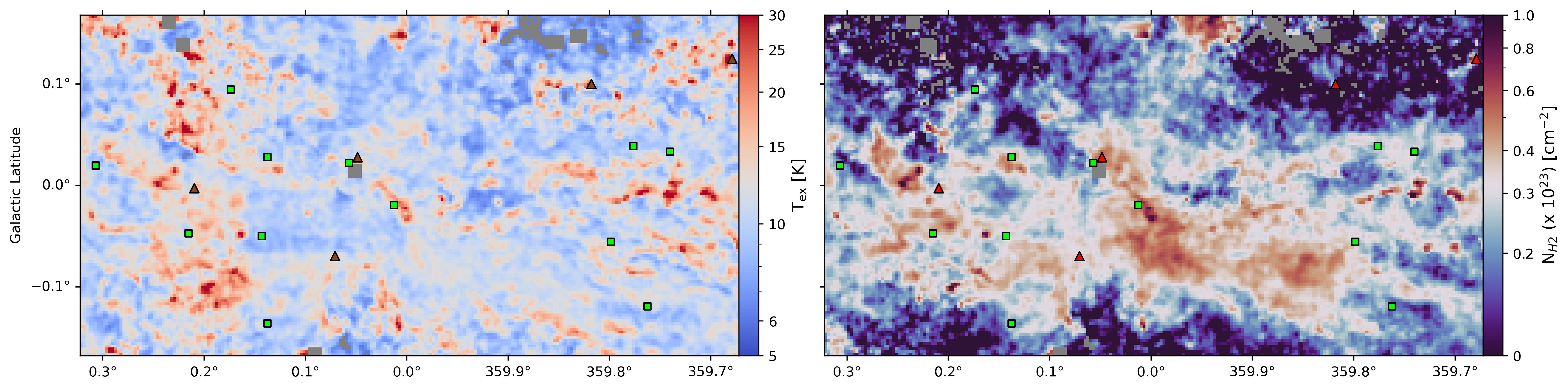}
    \includegraphics[width=2\columnwidth]{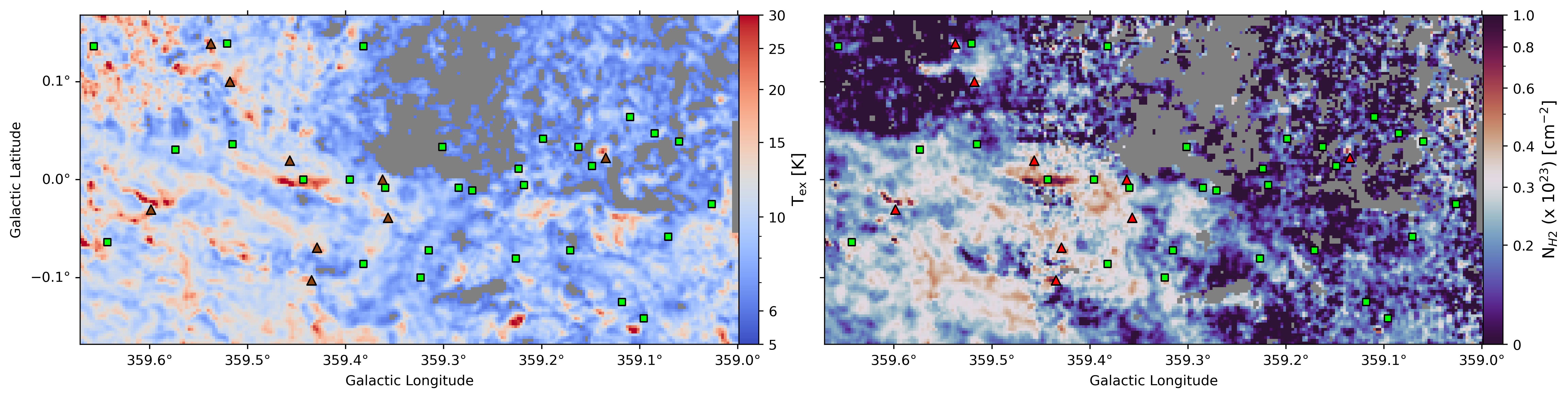}
    \caption{Comparison of the excitation temperature (\textbf{left}), and the H$_{2}$ column density (\textbf{right}) within Sgr\,A (\textbf{middle}), Sgr\,B (\textbf{top}) and Sgr\,C (\textbf{bottom}). Triangle markers show 70-$\micron$ Hi-GAL sources associated with a hotspot, while those without are shown in square markers. The temperature and column density are plotted on a logarithmic scale. This emphasizes the structural differences between them, such as those seen within the 20-km$\rm{s}^{-1}$ and 50-km$\rm{s}^{-1}$ clouds. Grey pixels represent regions where $\tau_r > 2.25$, thus no valid solution to Equation\,\ref{Eqn: brents} is found.}
    \label{fig:sgr_a_comparison}
\end{figure*}

Our estimated $^{13}$CO-traced molecular gas mass for the observed CMZ region is $M_{\rm{gas}} = (7^{+7}_{-3.5}) \times 10^{6}$\,M$_{\sun}$, based on the integrated H$_2$ column density over the inner $|\ell| < 1^{\circ}$ and $|b| < 0\fdg1$. While this is lower than some previous estimates for the full CMZ, the reduced mass is expected due to our spatial coverage.

For instance, \citet{Dahmen1998} estimated a total CMZ mass of $3 \times 10^{7}$\,M$_{\sun}$ from C$^{18}$O ($J=1\to0$) observations covering a broader region ($-0\fdg9 < |b| < 0\fdg75$). Our estimate is therefore about a factor of 4 smaller, consistent with the smaller observed area. In contrast, \citet{nagai2007} derived a mass of $2 \times 10^{7}$\,M$_{\sun}$ from a more confined region ($|b| < 0\fdg2$), still larger than our footprint but more comparable in latitude range.

Similarly, \citet{Battersby2025} reported a dense gas mass of $M_{\rm gas} = (2^{+2}_{-1}) \times 10^{7}$\,M$_{\sun}$ using {\it Herschel} dust-continuum observations above $N({\rm H}_2) > 10^{23}$\,cm$^{-2}$. To enable a direct comparison despite the different spatial coverage, we scaled the \citet{Battersby2025} column-density estimate to match our observed area, thus the remaining differences in our values are primarily due to methodology and tracer sensitivity rather than field-of-view effects. When scaled, their NH$_{2}$ column-density result produces an estimated H$_{2}$ mass of $(9^{+9}_{-4.5}) \times 10^{6}$\,M$_{\sun}$. While our estimate is lower in absolute terms, it remains consistent with previous measurements when accounting for differences in spatial coverage, resolution, and methodology. Therefore our result supports a robust, convergent picture of CMZ molecular gas mass falling in the range $(1-3) \times 10^{7}$\,M$_{\sun}$ across studies. A comparison of these mass estimates with their reported uncertainties can be see in Fig.\,\ref{fig:mass_comp}.

In addition to the total mass across the full field, we also determine the molecular gas mass within the three major CMZ complexes individually. Using the H$_2$ column density within each region, we obtain $M_{\rm gas} = 2.3 \times 10^{6}$\,M$_{\sun}$ for Sgr\,A, $3.2 \times 10^{6}$\,M$_{\sun}$ for Sgr\,B, and $1.7 \times 10^{6}$\,M$_{\sun}$ for Sgr\,C. These values reflect the well-known structural differences between the complexes: Sgr\,B dominates the mass budget, consistent with its extensive high-column-density gas and widespread star-forming material, whereas Sgr\,A and Sgr\,C each contribute smaller but still significant reservoirs. The relative proportions we recover shows that Sgr\,B contains roughly half of the total mass of the CMZ, with Sgr\,A and Sgr\,C contributing the remainder.

\begin{figure*}
    \centering
    \includegraphics[width=2\columnwidth]{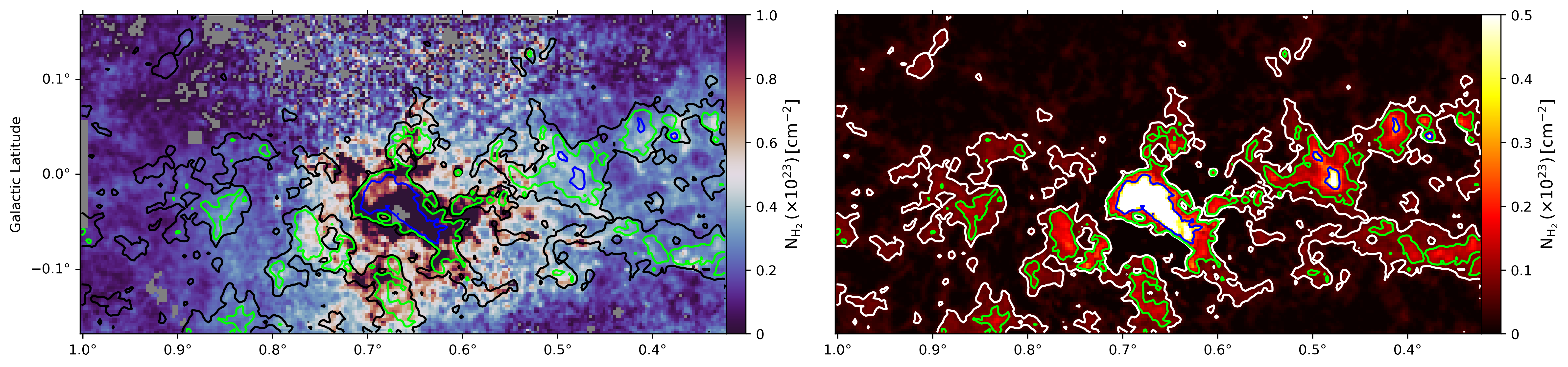}
    \includegraphics[width=2\columnwidth]{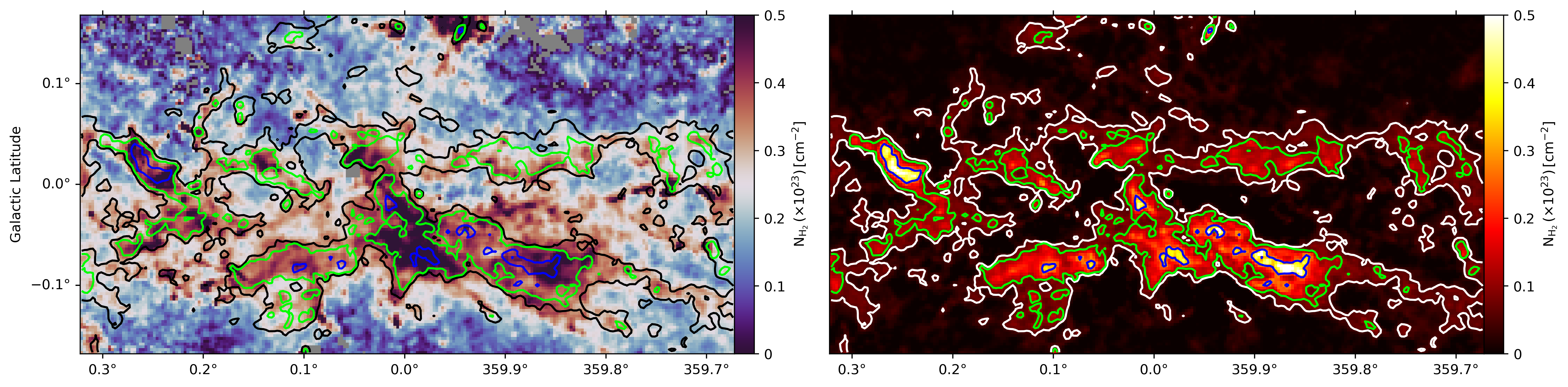}
    \includegraphics[width=2\columnwidth]{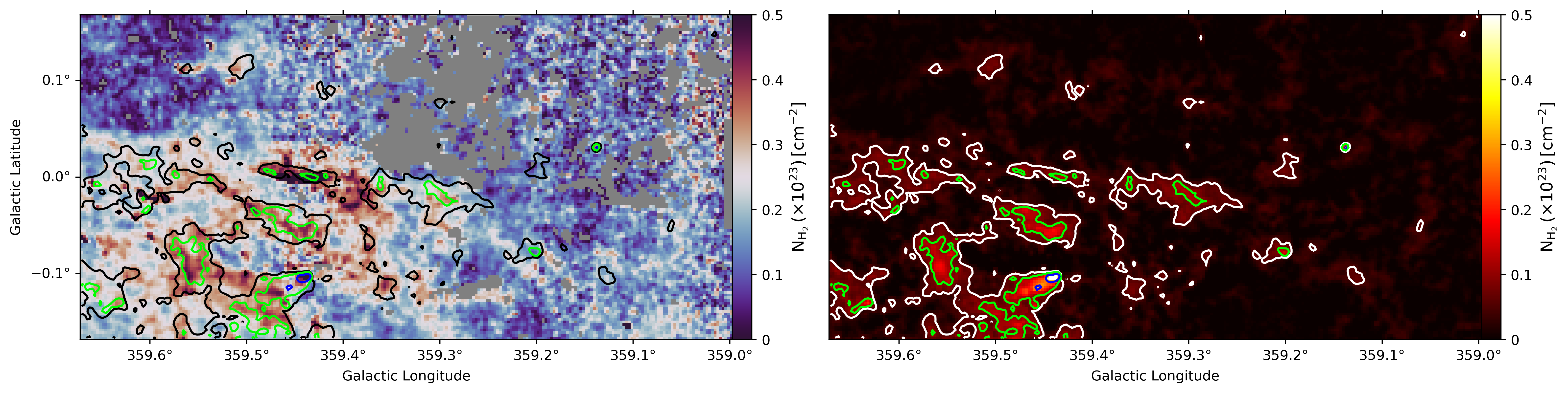}
    \caption{Comparison of the H$_{2}$ column density from this work (\textbf{left}) and the H$_{2}$ column density derived using the 850-$\micron$ thermal dust continuum flux density maps from SCUBA2 \citep{Parsons2018} (\textbf{right}), within the regions around Sgr\,A (\textbf{middle}), Sgr\,B (\textbf{top}) and Sgr\,C (\textbf{bottom}), as in Fig.\,\ref{fig:13CO_comparison}. Both data sets are contoured at dust-derived H$_2$ column densities of 0.05 (black/white), 0.1 (lime), 0.3 $\times 10^{23}$\,cm$^{-2}$ (blue).}
    \label{fig:sgr_dust_comparison}
\end{figure*}

\subsection{Discussion}
\subsubsection{Excitation Temperature}
We derived $T_{\rm ex}$ under the assumption of LTE, where the level populations of each transition are assumed to follow a Boltzmann distribution at a single temperature. Under LTE, the excitation temperature $T_{\rm ex}$ is expected to approach the gas kinetic temperature $T_{\rm kin}$ if the gas density is sufficiently high for thermalisation; here, we assume this to be the case, such that $T_{\rm ex} = T_{\rm kin} = T_{\rm gas}$. 

However, if the gas density is below the critical density required for collisions to dominate the excitation and the $^{13}$CO ($J = 3 \to 2$) transition is sub-thermally excited, then $T_{\rm ex}$ may underestimate the true $T_{\rm kin}$. Nevertheless, the excitation temperature reflects the relative populations in  energy levels $J=1$ to 3. Therefore, even in sub-thermal conditions, $T_{\rm ex}$ should still be a reliable tracer of relative variations in $T_{\rm kin}$.

The results in Fig.\,\ref{fig:Tex_map}
show significant temperature variations across the CMZ. The most apparent of these is in and around the Sgr\,B1/B2 complex where the highest excitation temperatures are observed; here $T_{\rm ex} > 50\,\mathrm{K}$. These high temperatures are likely to be driven by the ongoing intense active star formation and feedback processes in this region \citep{Schmiedeke2016, Ginsburg2018b}. 

Temperatures are relatively lower in the Dust Ridge--a dense filamentary structure of molecular clouds extending from The Brick (G0.253+0.016) \citep{Henshaw2019, immer2012b} towards the Sgr\,B1/B2 complex (G0.070-0.059), with $T_{\rm ex} > 20$\,K and several hot spots approaching $\sim 50\,$K. This contrasts with the dust temperature of $T_{\rm dust} < 15\, \mathrm{K}$ observed in \citet{Molinari2011}, as well as in \citet{Battersby2025}, which reports a similar dust temperature. It has been suggested that $T_{\rm gas}$ and $T_{\rm dust}$ are thermally uncoupled \citep{henshaw2023}, as might be expected at lower densities, and this could be responsible for the disparity between the two.

Across the wider CMZ region, the excitation temperature generally remains below $20\,\mathrm{K}$. In regions such as Sgr\,A, the 20-km\,s$^{-1}$ (G0.07–0.04) and 50-km\,s$^{-1}$ (G0.06–0.07) clouds, the temperatures are similar, with minimal variation between each distinct cloud, and are similar to those shown in the dust temperature maps of \cite{Molinari2011} and \cite{Battersby2025}. 

\paragraph*{Potential heating sources:}
To assess whether the hotspots seen in Fig.\,\ref{fig:Tex_map} and Fig.\,\ref{fig:sgr_a_comparison} are associated with identifiable heating sources, we compared the excitation temperature map with the spatial distribution of 70-$\micron$ sources from the Hi-GAL compact source catalogue \citep{Elia2021}. For this analysis, we assume all 70-$\micron$ sources lie at the same distance as the CMZ, consistent with the value adopted for total gas mass estimation in Section~\ref{subsec: column_density_method}.

We identified temperature peaks as local maxima in the excitation temperature map that also lie above the 80th percentile of the temperature distribution. A 70-$\micron$ source was considered associated with a temperature peak if it lay within one beam width (30 arcsec), corresponding to the angular resolution of the excitation temperature map.

Out of approximately 70 compact 70-$\micron$ sources in the region, only 17 ($\sim$25\%) were found within 30 arcseconds of a temperature peak after applying quality and masking criteria. These are highlighted as black/red circles in Fig.\,\ref{fig:sgr_a_comparison}. The lack of association for the majority of sources suggests that 70-$\micron$ sources are not the dominant drivers of the observed excitation temperature peaks, suggesting that other heating mechanisms, such as turbulent dissipation, shocks, or cosmic rays are likely to be contributing to the heating of gas.

\begin{figure*}
    \centering
    \includegraphics[width=2\columnwidth]{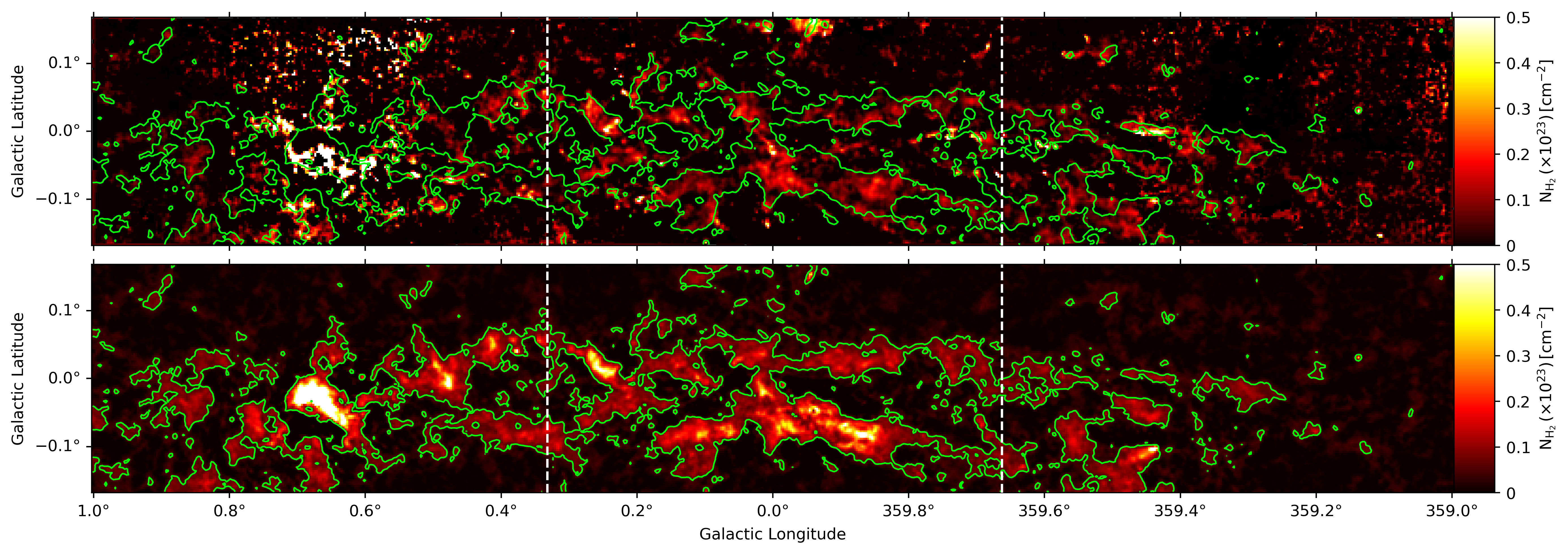}
    \caption{\textbf{Top}: The H$_{2}$ column density map shown in Fig.\,\ref{fig:H2_column_density} spatially filtered to remove structures larger than $\approx 8$ arcmin to match the SCUBA-2 spatial response, enabling a direct comparison between the two datasets. \textbf{Bottom}: The H$_{2}$ column density derived from the SCUBA-2 dust emission map. Both maps are overlaid with contours at 20 mJy\,arcsec$^{-2}$.}
    \label{fig:filtered_H2_comparison}
\end{figure*}

\paragraph*{Localised Gas Temperature Conditions in the CMZ:}
A more detailed look at the LTE-temperature structure can be seen in Fig.\,\ref{fig:sgr_a_comparison}. Here we present detailed $T_{\rm ex}$ (left column) maps across three sectors of the CMZ, each containing one of the Sagittarius complexes. 

The top panel (covering $0\fdg 32
< \ell < 1^{\circ}$) includes the Sgr\,B1/B2 complex and shows a highly complex temperature structure with pockets of gas that gradually become cooler outward from the region’s centre. The mean temperature in the area of the figure is $T_{\rm ex} = 14$\,K, although the temperature in Sgr\,B itself reaches well beyond $T_{\rm ex} = 120$\,K (see Fig.\,\ref{fig:sgr_a_comparison}). Much of the high-temperature structure is concentrated within $0\fdg55 < \ell < 0\fdg75$, with a narrow channel of low temperatures running horizontally through the structure  at $b \approx 0\fdg0$ which represents the separation of Sgr\,B2-North and Sgr\,B2-Main \citep{Schmiedeke2016}.

The middle panel of Fig.\,\ref{fig:sgr_a_comparison} covers the region $-0\fdg33 < \ell < 0\fdg32$ and includes Sgr\,A and the 20- and 50-km\,s$^{-1}$ clouds \citep{Kauffmann2017}. The temperature distribution is relatively uniform overall, but it contains a complex network of filamentary structures interspersed with compact high-temperature features. Some of these knots may correspond to embedded protostars or shock-heated regions driven by stellar feedback. While several hotspots coincide with 70-$\micron$ sources, many do not, suggesting that a variety of heating mechanisms may contribute to the observed temperature structures.

Two prominent shell-like features are visible at $\ell = 0\fdg2$, $b = -0\fdg1$ and $\ell = 0\fdg2$, $b = +0\fdg1$ in the middle panel of Fig.\,\ref{fig:sgr_a_comparison}. Both exhibit partial cavities with elevated excitation temperatures along their rims and also warmer gas near their centres. Such temperature structures are consistent with feedback-driven bubbles observed in other molecular environments, typically formed by OB cluster winds or protostellar outflows \citep{Arce2011, Beaumont2010, Barnes2017b}, and supported by radiative-hydrodynamic simulations \citep{Dale2012, Kim2023}.

The lower of these two features aligns closely with the well-known 'Sickle' H\,\textsc{ii} region (G0.18–0.04), which hosts the Quintuplet Cluster and is a canonical example of stellar feedback shaping the surrounding gas \citep{YusefZadeh1987, Lang1997, Wang2002}. The  morphology we observe here, a hot rim and centrally peaked $T_{\rm ex}$, is consistent with this interpretation.

The upper, bow-shaped shell is located near the end of the Arches filaments and coincides with the termination point of a prominent non-thermal filament visible in MeerKAT radio continuum \citep{Heywood2022}. While its temperature morphology resembles a feedback-driven cavity, with a curved, hot rim enclosing a lower-temperature core, there is no direct positional match with the Arches Cluster itself, nor with a known H\,\textsc{ii} region. Its alignment with the filament structure suggests a possible link to large-scale dynamic processes in the CMZ, but its origin remains uncertain without clearer evidence of internal heating sources.

In the immediate vicinity of Sgr\,A$^*$ ($\ell \approx 0\fdg0, b \approx 0\fdg0$), we see a localised elevation in excitation temperature, coinciding with a peak in H$_2$ column density (Fig.\,\ref{fig:sgr_a_comparison}, middle panel). However, this hotspot is not associated with any nearby 70-$\mu$m source within one beam width, suggesting that it may not be heated by an embedded protostar. Instead, the elevated temperature could reflect shock heating, external irradiation from nearby clusters, or other non-localised mechanisms.

The bottom-left panel of Fig.\,\ref{fig:sgr_a_comparison}, covering $359^{\circ} < \ell < 359\fdg67$, shows the Sgr\,C region. The mean temperature here is low at $\sim9$\,K and features a temperature distribution similar to Sgr\,A. Regions of cooler temperatures prominently surround warmer, denser cores, suggesting common physical conditions across these sectors. There are prominent networks of filamentary structures, with localised peaks in excitation temperature most of which are associated with a nearby 70–$\micron$ source.

\subsubsection{Column Density}
The right-hand panels of  Fig.\,\ref{fig:sgr_a_comparison} display the distribution of H$_{2}$ column density across the same three CMZ sectors discussed above.

In the top panel of Fig.\,\ref{fig:sgr_a_comparison}, covering the region around the Sgr\,B1/B2 complex, we observe a peak column density of $2 \times 10^{25}\,{\rm cm}^{-2}$ within Sgr\,B2 itself, and an average value of $\sim3 \times 10^{22}\,{\rm cm}^{-2}$ across the field.

The middle panel of Fig.\,\ref{fig:sgr_a_comparison} shows the region around Sgr\,A and the 20- and 50-km\,s$^{-1}$ clouds. Here, distinct high-column-density features are apparent, more clearly separated than in the corresponding excitation temperature map. The mean column density is $2 \times 10^{22}\,{\rm cm}^{-2}$, with a peak of $6 \times 10^{23}\,{\rm cm}^{-2}$. While temperature variations are more modest, some correspondence between warm gas and high column density is evident, suggesting that heating may be associated with dense structures. However, not all dense regions exhibit elevated temperatures, consistent with expectations that $^{13}$CO-traced gas can remain relatively cool. Near Sgr\,A$^*$ ($\ell \sim 0\fdg0$, $b \sim 0\fdg0$), a localised column density enhancement is observed, reflecting an accumulation of dense gas near the Galactic Centre, although there is no unique feature directly identifiable with Sgr\,A$^*$ itself.

In the bottom panel of Fig.\,\ref{fig:sgr_a_comparison}, covering the Sgr\,C region ($359\fdg67 < \ell < 359^{\circ}$), we observe column-density structures similar to those near Sgr\,A and Sgr\,B. Dense cores are embedded within a network of more diffuse filamentary gas, with peak column densities around $2 \times 10^{23}\,{\rm cm}^{-2}$.

\paragraph*{Gas structure Validation with SCUBA-2 Dust Maps:}

In order to investigate the validity of the small-scale structure in the molecular-gas results through quantitative morphological tests, we compare our $N({\rm H}_2)$ result to those derived from 850-$\micron$ thermal continuum dust emission data from the JCMT SCUBA-2 survey by \cite{Parsons2018}, assuming a constant dust temperature, $T_{\rm dust} = 20\,$K, and constant dust emissivity, $\beta = 2.0$ \cite{Parsons2018}. While the dust emission used to derive the total column density also traces the atomic gas phase, in the case of constant dust temperature and emissivity, and assuming the dust and gas are well mixed, the 850-$\micron$ flux density is proportional to $N({\rm H}_2)$. To convert the 850-$\micron$ emission into an H$_{2}$ column density we use the following formula.

\begin{equation}
N_{\rm H_2} = \frac{I_\nu}{\kappa_\nu \, B_\nu(T) \, \mu_{\rm H_2} \, m_{\rm H}}
\end{equation}

where $I_\nu$ is the observed 850-$\micron$ intensity, $\kappa_\nu$ is the dust opacity, $B_\nu(T)$ is the Planck function at dust temperature $T$, $\mu_{\rm H_2}$ and $m_{\rm H}$ are the mean molecular weight per hydrogen molecule, and mass of a hydrogen atom, given in section \ref{subsec: column_density_method}.

In Fig.\,\ref{fig:sgr_dust_comparison}, we compare the gas-derived $N({\rm H}_2)$ result to that derived from the 850-$\micron$ thermal continuum dust emission data. Statistical analysis and by-eye comparison suggest a good degree of correspondence between the molecular gas and dust emission in regions surrounding Sgr\,A and Sgr\,C, where both the general and some small-scale structure aligns well across the maps. Some variations in the finer details can be seen, which may be attributed to differences in excitation conditions and are unlikely to be noise-related. In the region containing Sgr\,B1/B2, a broad correspondence is observed, but the agreement in finer details is less pronounced, this poorer correspondence could be due to variations in dust temperature or excitation conditions that violate our assumption of constant $T_{\rm dust} = 20\,$K.

To test these subjective impressions quantitatively, we have computed the two-dimensional Kolmogorov–Smirnov (KS) test \citep{peacock1983} and the Spearman correlation between the full column-density map shown in Fig.\,\ref{fig:H2_column_density} and the corresponding SCUBA-2 map, subset regions of which are displayed in the right-hand column of Fig.\,\ref{fig:sgr_dust_comparison}.

The 2D KS test extends the traditional one-dimensional KS test by comparing the empirical cumulative distribution functions (ECDFs) of two samples in two dimensions. For each pixel $(x, y)$, the ECDF is defined as the fraction of the total data contributed by pixels with coordinates less than or equal to $(x, y)$, and the test statistic is the maximum absolute difference between the ECDFs of the two datasets over all such points. This provides a measure of how similar or different the two distributions are in a spatial sense. The p-value is then estimated using the survival function of the KS distribution.

The test returned a KS statistic of 0.31 and a p-value of 0.99. This suggests that the differences between the two distributions are not statistically significant, implying that the dust emission and H$_2$ column density are broadly similar in their overall distribution. 

The standard Spearman test was extended to two dimensions by separately calculating the pairwise correlation coefficients between the corresponding pixel values for the $x$ and $y$ axes of the H$_2$-column-density and SCUBA-2 maps. These were then combined using Fisher’s $z$-transformation \citep{Fisher1915} to obtain a single correlation coefficient that captures spatial dependencies in both directions. This method ensures that the test accounts for structure in the full two-dimensional distribution rather than treating each axis independently.

The resulting coefficient of $\rho = 0.54$ (p-value = $1.3\times10^{-44}$) indicates a highly significant, moderate correlation with considerable scatter. This implies that regions of higher CO-derived column density also tend to have higher 850-$\micron$ derived H$_{2}$ column-density, although the relationship is not strictly monotonic, possibly due to variations in dust temperature and emissivity (both of which were assumed constant here), or the presence of CO-dark H$_2$ gas. This may also be in part due to the SCUBA2 observing method and initial data reduction, the latter involving fitting to any extended background emission (which is mostly generated by the atmosphere) and removing it iteratively. This results in spatial filtering such that any emission with spatial frequencies greater than $\sim8$\,arcmin is suppressed; thus the comparison between our derived H$_{2}$ column-density map and the SCUBA2 map does not account for larger-scale structures.

\begin{table}
    \centering
    \caption{Statistical tests on high-pass filtered H$_{2}$ column-density maps shown in Fig.\,\ref{fig:filtered_H2_comparison} by region.}
    \label{tab:regional_tests}
    \begin{tabular}{lcccc}
        \hline
        \hline
        Region & KS Statistic & KS $p$-value & Spearman $\rho$ & $p$-value \\
        \hline
        Sgr~A & 0.62 & 0.83 & 0.50 & $7\times10^{-11}$ \\
        Sgr~B & 0.57 & 0.90 & 0.12 & 0.42 \\
        Sgr~C & 0.59 & 0.87 & $-0.09$ & $ 2\times10^{-29}$ \\
        \hline
        \textit{Full Map} & \textit{0.69} & \textit{0.73} & \textit{0.08} & \textit{0.10} \\
        \hline
        \hline
    \end{tabular}
\end{table}

To account for the spatial filtering inherent to the SCUBA-2 reduction, we repeated both statistical tests using the gas-derived H$_2$ column-density map after applying a Gaussian high-pass filter to remove all structure on scales larger than $\sim 8$\,arcmin (Fig.\,\ref{fig:filtered_H2_comparison}). This allowed for a more direct comparison of their small-scale structure.

The 2D KS test on the filtered maps produced a KS statistic of 0.69 ($p = 0.73$), indicating that the overall spatial \textit{distributions} of the two filtered datasets remain broadly similar, though with a larger maximum deviation between their CDFs.

In contrast, the two-dimensional Spearman correlation drops substantially once the large-scale emission is removed, yielding $\rho = 0.08$ with a non-significant $p = 0.1$. This suggests that the moderate correlation observed in the unfiltered maps is dominated by large-scale emission gradients. After matching the spatial response of the two datasets, the remaining small-scale morphology of the full map shows no statistically significant correlation.

Given this result, we performed the same filtered analysis separately on the Sgr~A, B, and C regions to determine if a particular region was driving the global signal. The results are summarised in Table\,\ref{tab:regional_tests}. The KS test confirms that the spatial distributions of the dust-derived and filtered gas-derived H$_{2}$ column-density maps are not significantly different in any region ($p > 0.83$ in all cases). However, the correlation behaviour varies dramatically:
\begin{itemize}
    \item The \textbf{Sgr\,A} region retains a moderate but significant correlation ($\rho = 0.50$, $p \ll 0.01$).
    \item The \textbf{Sgr\,B} region shows no significant correlation ($\rho = 0.12$, $p = 0.42$), consistent with the visual impression that its gas emission is dominated by large-scale structures not present in the SCUBA-2 data.
    \item The \textbf{Sgr\,C} region shows a significant but very weak \textit{anti}-correlation ($\rho = -0.09$, $p \ll 0.01$).
\end{itemize}

This regional decomposition reveals that the lack of a significant global correlation in the filtered maps is primarily due to the Sgr~B region, which dominates the total area. The strong correlation in Sgr~A is washed out in the whole-map analysis, indicating that the relationship between dust emission and gas tracers is not uniform across the Galactic Centre.

\section{Star formation efficiency in the CMZ} \label{sec: SFE}
\begin{figure*}
    \includegraphics[width=2\columnwidth]{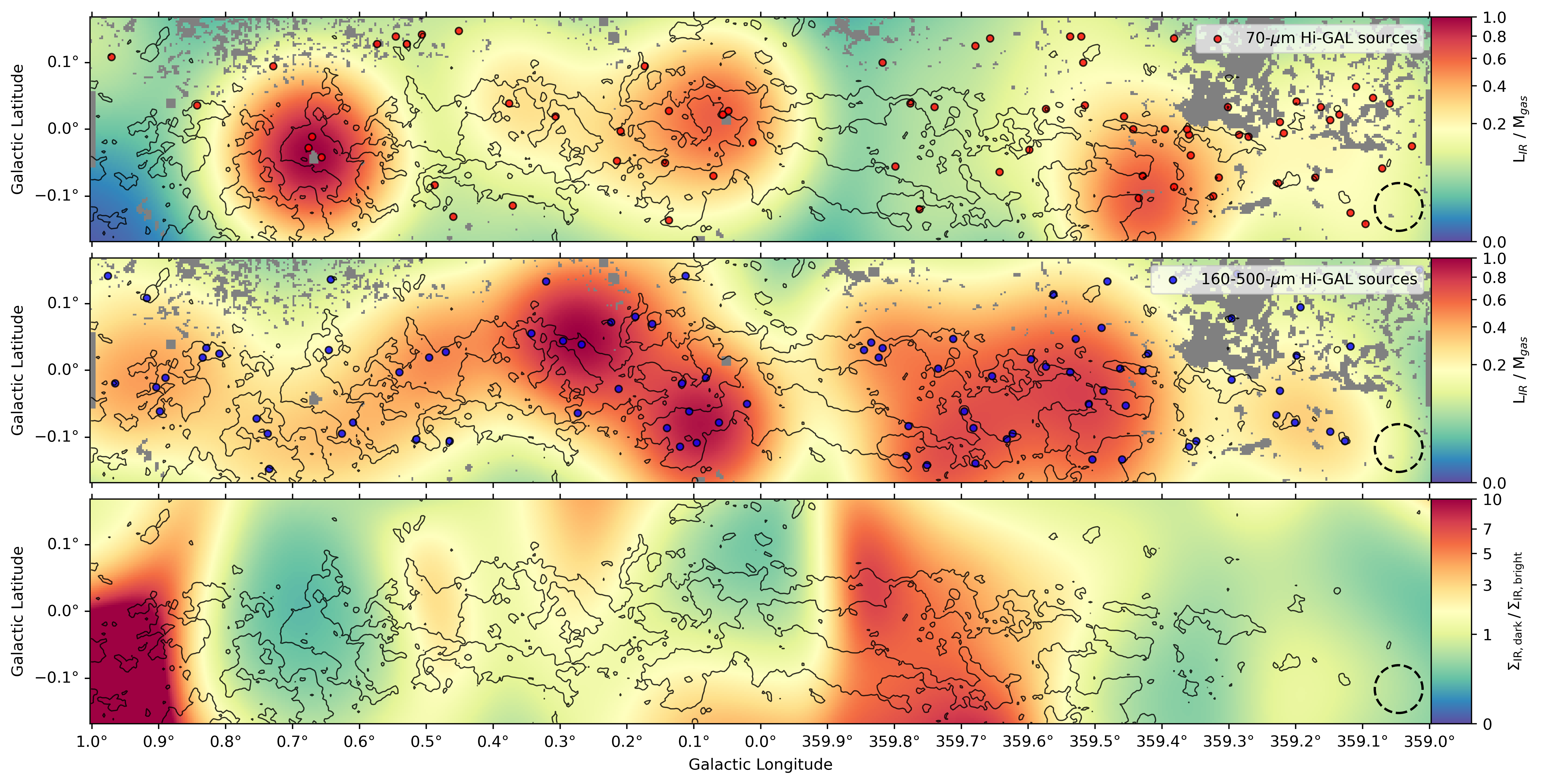}
    \caption{\textbf{Top}: The normalised star formation efficiency for the CMZ using 70-$\micron$ sources (red circles) from HiGAL \citep{HIGAL2016, Elia2021}. \textbf{Middle}: As for the top panel but showing the star formation efficiency using sources detected at 160--500-$\micron$ but not at 70-$\micron$ (blue circles). \textbf{Bottom}: The luminosity surface-density ratio, low values indicate regions nearing the end of their star forming phase. A luminosity surface-density ratio equal to 1 represents no significant relative change in SFE over time. The contours show the SCUBA2 dust emission at 20 (black) mJy\,arcsec$^{-2}$. The black circles in the bottom right represent the respective smoothing scale applied to the H$_2$ column-density maps. Grey pixels here represent values which returned a NaN, thus no valid solution to Equation \ref{Eqn:SFE} is found.}
    \label{fig:SFE_map}
\end{figure*}

In a previous work \citep{King2024}, we examined the relationship between the distribution of compact Hi-GAL sources and that of the molecular gas traced by $^{13}$CO in the CMZ.  We found that 70-$\micron$-bright sources, which are typically associated with actively star-forming regions, tend to be less spatially correlated with the dense molecular gas, suggesting that many of these sources may be located in foreground spiral arms rather than within the CMZ itself. In contrast, 160-500-$\micron$ sources without 70-$\micron$ detections, which may represent earlier stages of star formation, showed a stronger correlation with the $^{13}$CO spatial distribution, suggesting a closer association with the dense molecular gas in the CMZ. This gives us important  context for evaluating the star formation efficiency (SFE) across the CMZ.

The instantaneous SFE can be defined as the fraction of mass in dense CO-traced clouds that has formed stars over a given timescale \citep{Eden2015}. However, as the stellar mass cannot be measured directly, the luminosity of Young Stellar Objects (YSOs) embedded within these clouds serves as an indirect measure of their star formation activity. Therefore, the SFE can be parameterised by the ratio of the bolometric infrared (IR) luminosity of YSOs to the mass of molecular gas \citep{Urquhart2013a}, given by:

\begin{equation} 
    \label{Eqn:SFE}
    SFE = \frac{L_{\rm IR}}{M_{\rm gas}} = \frac{1}{M_{\rm gas}} \cdot \int_{0}^{t} \frac{dL}{dt}\,dt
\end{equation}

This is equivalent to the ratio of the infrared surface brightness ($\Sigma_{\rm IR}$) to the column density of molecular hydrogen ($N({\rm H_2})$), as long as the angular resolution or sampling scale is the same for both measurements. This ratio is independent of distance, meaning that the presence of different, often unreliable, distances for the clouds in the data does not affect our results, although overlapping line-of-sight features will be merged in the result because of this. 

It is important to note that we are not deriving absolute values of SFE, in terms of a fractional mass, but are using the $L_{\rm IR}/M_{\rm gas}$ parameter as a proxy to trace relative variations in SFE across different regions.

Given that the bolometric luminosity of infrared sources, $L_{\rm IR}$, does not scale linearly with the total stellar mass, the luminosity of young stellar objects (YSOs) is influenced not only by their mass but also by their evolutionary stage and accretion rate. In the early phases of star formation, the luminosity is dominated by  accretion processes, leading to significant variation among individual YSOs. For a fully populated initial mass function (IMF), the total luminosity of main-sequence stars follows a relation of $L \propto M^{-3.5}$ for individual stars, and $L \propto M^{2}$ for clusters \citep{Eden2015}. However, this relationship may be altered depending on whether the IMF is fully sampled. Here, we assume that it is, and that evolutionary timescales are short compared to the lifetime of a molecular cloud. We further assume that the infrared luminosity is primarily driven by high-mass main-sequence or near-main-sequence stars, which have already undergone most of their luminosity evolution. As a result, their brightness remains relatively stable during the IR-luminous phase \citep{Urquhart2022}.

The infrared luminosity, $L_{\rm IR}$, is derived from the integrated continuum fluxes of YSOs detected by Hi-GAL across multiple wavelengths, while the gas mass, $M_{\rm gas}$, is obtained from CO-traced observations of the molecular gas. These two quantities are independently measured and derived through distinct observational techniques, and thus exhibit no inherent correlation beyond their physical connection via the star formation process. The ratio $L_{\rm IR}/M_{\rm gas}$ can inherently reflect evolutionary effects, particularly when evaluated for individual sources or when both $L$ and $M$ are derived from the same continuum measurements. However, we argue that the 70-$\micron$-bright and -dark phases (those traced by 160–500-$\micron$ sources) are expected to be relatively short-lived compared to overall cloud lifetimes and are here averaged over statistically large samples, therefore the resulting $L_{\rm IR}/M_{\rm gas}$ maps robustly reflect the instantaneous star-formation efficiency in each phase, and comparison between the phases provides information on evolution.

\vspace{-0.5cm}

\subsection{Method}\label{subsec: SFE_method}
We have adopted a surface-density based approach rather than segmenting the CMZ into discrete molecular clouds. This method allows us to trace large-scale spatial variations in SFE across the CMZ without imposing an artificial division into clouds, which can be strongly dependent on the choice of cloud identification algorithm and input parameter values. By correlating the infrared luminosity surface density produced by compact sources with gas mass surface density on a pixel-by-pixel basis, we are able to probe the variations in  $L_{\rm IR}/M_{\rm gas}$ throughout the CMZ.

We conducted our analysis using all five Hi-GAL wavebands 
(70, 160, 250, 350 and 500\,$\micron$) from the \citet{Elia2021} compact-source catalogue. 70-$\micron$ emission serves as a tracer of active star formation within dense cores, while the longer-wavelength submillimetre bands (160 to 500-$\micron$) probe dense, cold structures associated with embedded star-forming regions. Although bolometric luminosities for Hi-GAL clumps are available in the \citet{Elia2021} catalogue, we opted to compute luminosities using only these five Hi-GAL wavelengths, excluding shorter-wavelength fluxes (21–24-$\micron$) that may be dominated by transiently heated dust, as well as longer wavelengths (870–1100\,$\micron$) that may trace more extended cold dust unrelated to compact dense cores.

The SFE analysis focused on the ratio of the bolometric infrared luminosity ($L_{\rm IR}$) of Hi-GAL sources to the gas mass estimated using Equation\,\ref{Eqn: H2_mass_equation} at each pixel in H$_2$ column density in Fig.\,\ref{fig:H2_column_density}. $L_{\rm IR}$ is given by:

\begin{equation}
    L_{\text{IR}} = 4\pi\,d^2\int_{\nu_{\min}}^{\nu_{\max}} S_{\nu} \, d\nu.
\end{equation}

where $S_{\nu}$ is the flux density, $\nu_{\min}$ and $\nu_{\max}$ are the Hi-GAL wavelengths converted to the frequency domain and $d$ is the distance. In practice we only require the integrated flux, since the ratio of this to the mass column density gives us $L_{\rm IR}/M_{\rm gas}$ without the need to define actual source distances. We employed the trapezium method to numerically approximate the integral.

To calculate a continuous bolometric luminosity surface density distribution from the discrete Hi-GAL sources, we used a Gaussian Kernel Density Estimation (KDE). The smoothing scale was determined using Silverman’s rule of thumb for bandwidth estimation \citep{silverman1986}, given by:

\begin{equation} \label{Eqn: silverman}
h = \left( \frac{4 \langle \sigma \rangle^5}{3n} \right)^{1/5},
\end{equation}

where $\langle \sigma \rangle$ is the standard deviation of the source coordinates along one spatial axis, and $n$ is the number of data points (i.e., the number of 70-$\micron$ and 160–500-$\micron$ Hi-GAL compact sources). While Silverman's rule is traditionally applied by averaging the spread across both axes (assuming isotropy), in our case the spatial distribution of sources is highly anisotropic, with a significantly narrower spread along Galactic latitude. We therefore adopted a more representative, symmetric smoothing scale by applying twice the bandwidth estimated from the latitude axis alone ($2h_y$), thereby avoiding excessive smoothing along the more extended longitude direction, and to avoid zeros in the resulting interpolation.

This smoothing scale (equivalent to $2h_y$) was applied to the H$_{2}$ column-density map (Fig.,\ref{fig:H2_column_density}) to match the effective resolution of the surface-luminosity maps (see lower two panels of Fig.,\ref{fig:Surface_luminosity}). As this smoothing scale is coarser than the native resolution of the H$_2$ map, small-scale variations in SFE are not resolved. In any case, there is a minimum scale below which the concept of SFE becomes meaningless. As such, our analysis focuses on the larger-scale structure of the star formation efficiency distribution.

To produce the SFE maps shown in Fig.\,\ref{fig:SFE_map}, the Hi-GAL surface luminosity was divided by the smoothed H$_{2}$ column density to produce two distinct SFE distributions: one for actively star-forming regions \citep{Moore2012, Eden2015}, and another for those apparently at earlier stages of evolution \citep{Eden2012, Eden2013}.

\vspace{-5mm}

\subsection{Results}
Fig.\,\ref{fig:SFE_map} shows the distribution of the SFE, defined as $L_{\rm IR} / M_{\rm gas}$, across the CMZ. The values are normalised to a common scale to emphasise relative variation rather than absolute efficiency. The top panel traces the SFE derived from 70-$\micron$-bright sources, representing a snapshot of current star formation activity, while the bottom panel shows the instantaneous SFE based on 160–500-$\micron$ sources, indicative of potential future star formation assuming that these regions will eventually form stars.

In the top panel, the 70-$\micron$-bright SFE is enhanced in three well-known regions. We observe elevated SFE in the Sgr\,B1/B2 complex (G0.070-0.059), consistent with its well established role as one of the most active star-forming regions in the Galaxy. Sgr\,B2 in particular has been the subject of extensive study due to its exceptional mass, density, and ongoing high-mass star formation \citep[e.g.,][]{Ott2014, Ginsburg2018a, Ginsburg2018b, Mills2018, Schwoerer2019, Santa-Maria2021}. Elevated SFE is also seen in the vicinity of the Arches cluster (G0.121+0.017) and in Sgr\,C (G-0.49-0.130), although not to the same extent as in Sgr\,B1/B2. These enhancements are consistent with previous studies identifying these regions as significant sites of star formation in the CMZ \citep[e.g.,][]{Figer2002, Kendrew2013}.

Slightly elevated SFE is also observed across the dust ridge clouds (G0.20+0.003), particularly to the left of the Arches cluster, with the exception of clouds e\&f (G0.490+0.008), which display significantly lower SFE compared to the other dust ridge clouds (clouds b, c, d, and the Brick). This forms a notable boundary of low SFE between the dust ridge and the Sgr\,B1/B2 complex.

The middle panel shows a contrasting distribution. The 160--500-$\micron$-derived SFE appears elevated across much of the CMZ, but is notably lower toward the Sgr\,B1/B2 complex compared to the 70-$\micron$-bright SFE in the top panel. This reduction was initially attributed to saturation in the 160- and 250-$\micron$ PACS bands, which are known to saturate in high surface-brightness regions such as Sgr\,B2 \citep{Poglitsch2010, Griffin2010}. However, since both subsets of the surface luminosity used to derive the SFE (top and middle panels) include the 160- and 250-$\micron$ bands, any impact from saturation is present in both maps. Whereas, the 350- and 500-$\micron$ SPIRE bands are not significantly affected and continue to trace cold, extended emission.

Therefore, the observed drop in SFE in the middle panel is not an artifact of saturation, but probably reflects a genuine decrease in the number or luminosity of cold embedded Hi-GAL sources in Sgr\,B2. This suggests a lower star-formation efficiency in the future. This interpretation is further supported by the surface-luminosity ratio map in the lower panel of Fig.\,\ref{fig:SFE_map}, which highlights regions where current star formation is elevated relative to cold dust reservoirs. Regions such as Sgr\,B2 with high surface-luminosity ratios are consistent with being towards the end of their star-forming lifetimes. This is further supported by the surface-luminosity ratio map in the lower panel of Fig.\,\ref{fig:SFE_map}. Since this ratio is proportional the ratio of the two SFE values, it highlights regions where current SFE is reduced relative to the available gas reservoir. Regions such as Sgr\,B2 with low ratios are therefore consistent with being toward the end of their star-forming lifetimes.

The SFE across the dust ridge clouds is significantly elevated in this map compared to the 70-$\micron$-bright counterpart, with the Brick and the Three Little Pigs clouds (G0.110-0.079) now showing comparably high SFEs. SFE also remains high toward Sgr\,C, as in the 70-$\micron$ map, but is now more spatially extended, including the Wiggles feature \citep{Henshaw2020} and surrounding clouds.

\vspace{-5mm}

\subsection{Discussion}
The $L_{\rm IR} / M_{\rm gas}$ values in Fig.\,\ref{fig:SFE_map} are shaped by two main factors: the star formation rate (SFR) per unit gas mass, integrated over a relevant timescale, and the luminosity function (LF) of embedded young stellar sources. In this study, the timescale is constrained by the evolutionary phases traced by Hi-GAL: 70-$\micron$-bright sources associated with protostars ($<5\times10^5$\,yr; \citealt{Mottram2011}) and cold dust structures or prestellar cores detected only at longer wavelengths (typically $\sim3\times10^5$\,yr; \citealt{Kirk2005}). These timescales are significantly shorter than the typical lifetimes of molecular clouds ($\sim10^7$–$5\times10^7$\,yr; \citealt{Jeffreson2018}), so $L_{\rm IR}/M_{\rm gas}$ effectively traces a snapshot of the current or near-future star formation activity.

The two SFE maps in Fig.\,\ref{fig:SFE_map} capture complementary stages in the star formation cycle. The 70-$\micron$-bright sources trace ongoing, embedded star formation, while the sources detected only at longer wavelengths (160–500\,$\micron$) represent colder, denser structures that have not yet formed protostars such as pre-stellar cores, filaments, or clumps. Their associated instantaneous SFE thus represents an incipient or potential future star formation efficiency.

A significant feature of Fig.\,\ref{fig:SFE_map} is the spatial contrast between the upper two maps. The bottom panel provides insight into the relative evolutionary potential of the SFE through the ratio of the luminosities shown in Fig.\,\ref{fig:Surface_luminosity}. A value of 1 represents no potential change in SFE over time, thus regions such as Sgr\,B1/B2, are likely to be at the end of their star-forming phase.  A higher value represents regions with an elevated potential for future star formation, such as the dust ridge clouds and Sgr\,C, which exhibit modest SFE distributions in the 70-$\micron$-bright map but show significantly higher values in the 160–500\,$\micron$ map. Conversely, areas with elevated current star formation — particularly Sgr\,B1/B2 -show a suppressed incipient SFE. This relative anti-correlation suggests a spatial variation in evolutionary stage: while some regions are currently active, others may be on the verge of entering a more intense star-forming phase.

Despite hosting a large gas reservoir and known luminous sources, the Sgr\,B1/B2 complex exhibits relatively low SFE in the 160–500\,$\micron$ map. This may reflect the combined effects of turbulent cloud kinematics \citep{Henshaw2016a} and elevated gas temperatures \citep{Ginsburg2016}, both of which can inhibit the collapse of dense structures. Alternatively, it may indicate that this region is transitioning out of a star-forming phase and entering a more quiescent evolutionary state.

Together, these findings suggest that while the CMZ exhibits localised sites of intense star formation (e.g., Sgr\,B2, Arches), it also harbours a widespread reservoir of cold, dense gas with the potential for future activity. This is quantitatively supported by the higher average incipient SFE in the 160-500-$\micron$ derived map ($SFE = 0.35$) compared to the current SFE shown in the 70-$\micron$ bright map ($SFE = 0.19$), suggesting we may be observing during a low-activity period preceding more widespread star formation.

The elevated incipient SFE across the dust ridge clouds, Sgr\,C, and in the Sgr\,A region shown in the lower panel of Fig.\,\ref{fig:SFE_map} implies that a broader starburst phase may be imminent across the CMZ and we may be observing during a low-activity period.

\paragraph*{Source distance considerations:} We assume that all sources presented here are physically located within the CMZ. Although precise distance measurements are lacking, the strong concentration of sources in Galactic latitude, combined with the known spread of foreground spiral arm emission in the CHIMPS2 data \citep{Eden2020, King2024}, supports this assumption. A small fraction of contamination from foreground structures cannot be ruled out, but is unlikely to significantly affect the overall trends discussed here.

\vspace{-0.5em}

\section{Summary} \label{sec: Summary}
We present an analysis of molecular gas properties and star formation activity in the CMZ, revealing several key findings:

The LTE gas temperature map across the CMZ shows a mean temperature of $12\pm 6$\,K and a median of $11^{+2}_{-2}$\,K, with the highest temperatures exceeding 30\,K observed around Sgr\,B (G0.68-0.05), and $T_{\rm ex} > 120$\,K within Sgr\,B. In contrast, gas temperatures near Sgr\,A appear significantly cooler with little variation in temperature among individual clouds, though overall the temperatures traced by $^{13}$CO appear consistent with dust temperatures presented in \cite{Molinari2011}. Excitation temperatures reveal a significant and extended filamentary structure surrounding Sgr\,A clouds, and in negative longitudes towards Sgr\,C.

Our column density estimates reveal a median $^{13}$CO column density of $(3^{+1}_{-1}) \times 10^{16}$\,cm$^{-2}$, corresponding to an H$_{2}$ column density of $(2^{+0.9}_{-0.7}) \times 10^{22}$\,cm$^{-2}$. The highest column density, $2 \times 10^{25}$\,cm$^{-2}$, is also found near Sgr\,B2, which coincides with its status as a prominent site of star formation. The total $^{13}$CO-traced mass in the CMZ is estimated to be $M_{\rm gas} = (7^{+7}_{-3.5}) \times 10^{6}\,M_{\sun}$, the upper limit of our estimate is consistent with \cite{Dahmen1998}, while our total estimate is consistent with \cite{nagai2007} and \cite{Battersby2025}. These agreements suggest that the total mass of the CMZ is well-established within these limits.

Mapping the instantaneous star formation efficiency (SFE) using Hi-GAL sources reveals clear spatial variations that distinguish between current and potential future star-forming regions in the CMZ. In the 70-$\micron$-bright map, which traces embedded, actively star-forming sources, the SFE is enhanced in several known regions, including the Sgr\,B1/B2 complex, the Arches cluster, and Sgr\,C. Modest enhancement is also seen across the dust ridge clouds. However, the overall distribution of SFE in dust ridge clouds remains relatively low compared to other regions such as the nearby Arches cluster.

In contrast, the SFE traced by 160–500-$\micron$ sources, which probe colder dust and prestellar structures shows a much broader and more uniform enhancement across the CMZ. This includes high values in the Brick, the Three Little Pigs clouds, the Arches region, and throughout both high and low latitudes, particularly at negative longitudes towards the Sgr\,C region. These features suggest the presence of widespread, dense gas that is potentially on the verge of forming stars. Interestingly, Sgr\,B1/B2 shows relatively low SFE in 160–500-$\micron$ tracers, despite its substantial gas reservoir. This may reflect physical conditions, such as turbulent kinematics and high gas temperatures—that inhibit gravitational collapse, or it may indicate that the region is evolving into a more quiescent phase.

Together, these maps support a dynamic evolutionary scenario for star formation in the CMZ. The widespread elevation of 160–500-$\micron$ SFE across quiescent clouds and inflow regions suggests that the CMZ is primed for a future increase in star formation activity, consistent with models of orbital inflow and cloud evolution. The contrasting morphologies between the two maps shows a clear progression from dense, cold gas to luminous protostellar activity, linking the observed SFE distribution to the larger dynamical cycle of gas inflow, accumulation, and star formation in the Galactic Centre.

\vspace{-7mm}

\section*{Acknowledgements}
The James Clerk Maxwell Telescope is operated by the East Asian Observatory on behalf of The National Astronomical Observatory of Japan; Academia Sinica Institute of Astronomy and Astrophysics; the Korea Astronomy and Space Science Institute; the Operation, Maintenance and Upgrading Fund for Astronomical Telescopes and Facility Instruments, budgeted from the Ministry of Finance (MOF) of China and administrated by the Chinese Academy of Sciences (CAS). Additional funding support is provided by the Science and Technology Facilities Council of the United Kingdom and participating universities in the United Kingdom and Canada. The Starlink software \citep{Currie2014} is currently supported by the East Asian Observatory. This research has made use of NASA’s Astrophysics Data System.

This publication is based in part on data acquired with the Atacama Pathfinder Experiment (APEX) under programmes 092.F-9315 and 193.C-0584. APEX is a collaboration among the Max-Planck-Institut fur Radioastronomie, the European Southern Observatory, and the Onsala Space Observatory.

\vspace{-7mm}

\section*{Data Availability} 
The reduced CHIMPS2 $^{13}\text{CO}$ CMZ data are available to download from the \href{https://www.canfar.net/storage/vault/list/CHIMPS2}{CANFAR archive}. The data are available as mosaics, roughly 2$^{\circ} \times 1^{\circ}$ in size, as well as individual observations. The data are presented in FITS format.

The processed SEDIGISM data products are available from the \href{https://sedigism.mpifr-bonn.mpg.de/index.html}{SEDIGISM database}, which was constructed by James Urquhart and hosted by the Max Planck Institute for Radio Astronomy.

\vspace{-0.5cm}

\bibliographystyle{mnras}
\bibliography{Paper_draft}

\appendix
\section{Smoothed Hi-GAL Surface Luminosity and H\texorpdfstring{$_{2}$}{2} column-density plots}
\begin{figure*}
     \includegraphics[width=2\columnwidth]{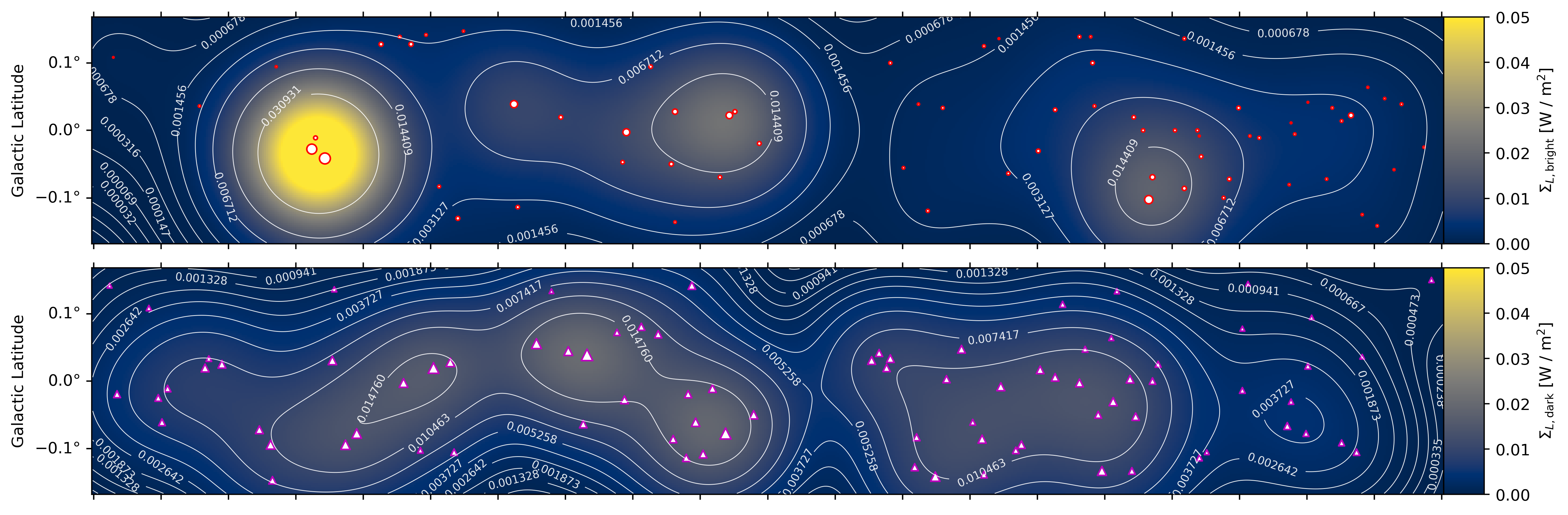}
     \includegraphics[width=2\columnwidth]{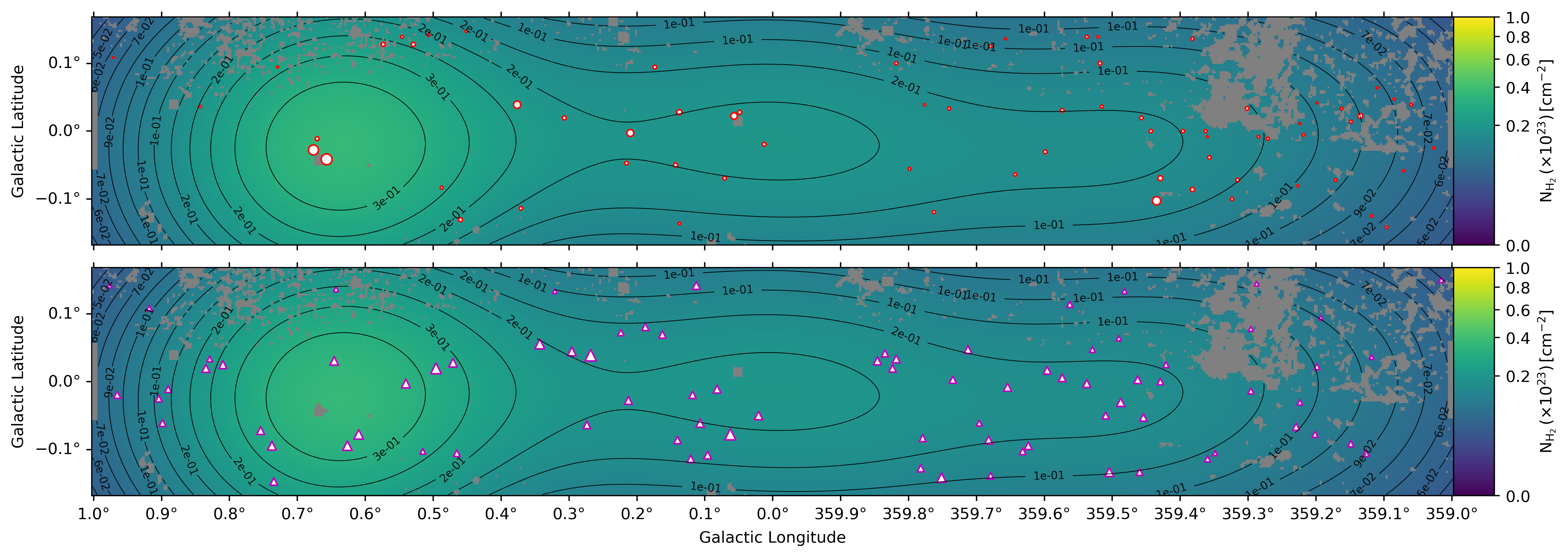}
\caption{
\textbf{Top}: The surface luminosity distribution of the 70-$\micron$ Hi-GAL sources used to determine the star formation efficiency (SFE) in Fig.\,\ref{fig:SFE_map}. 
\textbf{Second}: The surface luminosity distribution for the 160–500-$\micron$ sources. Each marker size is proportional to the square root of the source's total flux. Contours represent the surface luminosity, ranging from the minimum to maximum observed values, with labels indicating specific luminosity levels. \textbf{Third}: The H$_{2}$ column-density distribution smoothed to 8 arcmins. Each maker shows the position of 70-$\micron$ Hi-GAL sources as per the top panel. \textbf{Bottom}: As for third top but showing the 160-500-$\micron$ sources.} 
\label{fig:Surface_luminosity}
\end{figure*}

The top two panels of Fig.\,\ref{fig:Surface_luminosity} show the smoothed surface luminosity distribution of Hi-GAL sources, used to derive the luminosity-to-mass (L$_{\rm IR}$/M$_{\rm gas}$) in the main SFE analysis. The map shows the spatially smoothed luminosity surface density of compact sources detected at 70\,$\micron$ (top panel) and 160--500\,$\micron$ (second panel).

The lower two panels of Fig.\,\ref{fig:Surface_luminosity} show the H$_{2}$ column-density map (see Fig.\,\ref{fig:H2_column_density}) smoothed to the 8.5 arcmin resolution of the surface luminosity map. These maps serve as the basis for calculating the SFE maps shown in Fig.~\ref{fig:SFE_map}.

\bsp	
\label{lastpage}

\end{document}